\documentclass[aps,prl,reprint,showpacs,nofootinbib,superscriptaddress]{revtex4-1}
\usepackage{graphicx,comment}
\usepackage{amsfonts}
\usepackage{amsmath,amssymb}
\usepackage{nth}
\usepackage{bm}
\usepackage{color}
\usepackage{tikz}
\usepackage{pgfplots}
\usepackage{qcircuit}
\usepackage{float}
\pgfplotsset{
table/search path={csv/},
}
\usetikzlibrary{pgfplots.groupplots}
\usepackage[font=small,
   justification=justified,
   format=plain]{caption} 
\usepackage{subfig}
\newcommand\blfootnote[1]{%
  \begingroup
  \renewcommand\thefootnote{}\footnote{#1}%
  \addtocounter{footnote}{-1}%
  \endgroup
}

\usepackage{environ}
\NewEnviron{myhideenv}{%
  \setbox0\vbox{%
   \let\xxwrite\write
   \protected\def\write{\immediate\xxwrite}%
   {\tiny XX\BODY XX}}
  }

\usepackage{verbatim}
\usepackage{hyperref}
\usepackage{multirow}
\usepackage{tabularx}
\usepackage[normalem]{ulem}

\hypersetup{
     colorlinks   = true,
    citecolor    = blue
}
\usepackage{lineno}
\usepackage[thicklines]{cancel}
\usepackage{color, colortbl}
\definecolor{LightCyan}{rgb}{0.88,1,1}
\usepackage{url}
\usepackage{xurl}
    
{\rm }

\def\be{\begin{equation}}
 \def\ee{\end{equation}}
 \def\bea{\begin{eqnarray}}
 \def\eea{\end{eqnarray}}

\def\2{\frac{1}{2}}
\def\4{\frac{1}{4}}

\newcommand{\ket}[1]{\left| #1 \right\rangle}

\definecolor{red}{rgb}{1,0.,0}

\begin{document}

\title{Performant near-term quantum combinatorial optimization}
\author{Titus Morris}
\affiliation{Quantum Information Science Section, Oak Ridge National Laboratory, Oak Ridge, TN 37831, USA}

\author{Ananth Kaushik}
\affiliation{IonQ Inc, 4505 Campus Dr, College Park, MD 20740, USA}

\author{Martin Roetteler}
\affiliation{IonQ Inc, 4505 Campus Dr, College Park, MD 20740, USA}

\author{Phillip C. Lotshaw}
\affiliation{Quantum Information Science Section, Oak Ridge National Laboratory, Oak Ridge, TN 37831, USA}
\blfootnote{This manuscript has been authored by UT-Battelle, LLC under Contract No. DE-AC05-00OR22725 with the U.S. Department of Energy. The United States Government retains and the publisher, by accepting the article for publication, acknowledges that the United States Government retains a non-exclusive, paid-up, irrevocable, world-wide license to publish or reproduce the published form of this manuscript, or allow others to do so, for United States Government purposes. The Department of Energy will provide public access to these results of federally sponsored research in accordance with the DOE Public Access Plan (http://energy.gov/downloads/doe-public-access-plan).}

\date{\today}

\begin{abstract}
Combinatorial optimization is a promising application for near-term quantum computers, however, identifying performant algorithms suited to noisy quantum hardware remains as an important goal to potentially realizing quantum computational advantages. To address this  we present a variational quantum algorithm for solving combinatorial optimization problems with linear-depth circuits.  Our algorithm uses an ansatz composed of Hamiltonian generators designed to control each term in the target combinatorial function, along with parameter updates following a modified version of quantum imaginary time evolution. We evaluate this ansatz in numerical simulations that target solutions to the MAXCUT problem. The state evolution is shown to closely mimic imaginary time evolution, and its optimal-solution convergence is further improved using adaptive transformations of the classical Hamiltonian spectrum.  With these innovations, the algorithm consistently converges to optimal solutions, with interesting highly-entangled dynamics along the way. We further demonstrate the success of this approach by performing optimization of a truncated version of our ansatz with up to 32 qubits in a trapped-ion quantum computer, using measurements from the quantum computer to train the ansatz and prepare optimal solutions with high fidelity in the majority of cases we consider. The success of these large-scale quantum circuit optimizations, in the presence of realistic hardware constraints and without error mitigation, mark significant progress on the path towards solving combinatorial problems using quantum computers. We conclude our performant and resource-minimal approach is a promising candidate for potential quantum computational advantages.
\end{abstract}

\maketitle

\section*{Introduction}

Combinatorial optimization has long been viewed as a scientific computing challenge where quantum computers may have a significant impact \cite{farhi2001quantum,albash2018adiabatic,herman2023quantum,preskill2018quantum,abbas2023quantum}. In addition to their fundamental interest as examples of the NP-hard computational complexity class, combinatorial problems also have diverse applications in fields such as finance, logistics, and operations research.
However, the complexity of these problems is a major challenge to identifying high-quality solutions.  According to computational complexity theory, NP-hard instances cannot be solved exactly when using computational resources that scale polynomially with the problem size.
This has led to the widespread use of heuristic algorithms that yield approximate solutions in a reasonable runtime.
Improving these heuristics is an important topic of ongoing research, as even small improvements have the potential for significant impacts when reducing costs or improving efficiency at scale \cite{abbas2023quantum}.   

Quantum computers are rapidly improving and have been demonstrated to solve certain classically intractable problems \cite{arute2019quantum,decross2024computational,kim2023evidence,robledo2024chemistry}, motivating intensive research into understanding their potential utility for solving useful and computationally difficult problems.  Recent research into quantum combinatorial optimization algorithms has mainly focused on the quantum approximate optimization algorithm (QAOA) \cite{farhi2014quantum,hogg2000quantum} as well as its variants \cite{hadfield2019quantum,herman2023constrained,ushijima2021multilevel,tate2023warm,bartschi2020grover,wurtz2021classically,herrman2022multi,chandarana2022digitized,zhu2022adaptive,blekos2023review,goldstein2024convergence}. A variety of encouraging results have been obtained for QAOA in theoretical \cite{farhi2022quantum,wurtz2021MAXCUT,diez2023quantum,wurtz2022counterdiabaticity} and numerical analyses \cite{zhou2020quantum,lotshaw2021empirical,herrman2021impact,golden2023numerical,crooks2018performance,shaydulin2023parameter}, including arguments that QAOA outperforms conventional algorithms for ``large-girth" MaxCut instances \cite{basso2021quantum} and obtains a scaling advantage over conventional methods for the low autocorrelation binary sequences problem \cite{shaydulin2023evidence}. 

Despite encouraging progress, there are serious drawbacks to existing approaches such as QAOA. QAOA is based on a set of circuit operations which can be repeated indefinitely to improve its theoretical performance, with high performance typically requiring deep circuits containing many  layers \cite{basso2021quantum,shaydulin2023evidence,farhi2022quantum,lykov2023sampling,lotshaw2023approximate,marwaha2021local,farhi2020quantum,farhi2020quantum2}. However, deep circuits are incompatible with limitations of existing quantum computers, where noise severely impacts their theoretical and experimental performance as depth increases \cite{stilck2021limitations,lotshaw2022scaling,weidenfeller2022scaling,de2023limitations}. There have been a variety of important hardware demonstrations of QAOA and related approaches, but so far these have been limited to small numbers of layers, or small numbers of qubits, or both, and often with limited optimal solution probability \cite{ebadi2022quantum,maciejewski2023design,harrigan2021quantum,pagano2020quantum,niroula2022constrained,shaydulin2023qaoa,pelofske2023quantum,dupont2024quantum,maciejewski2024multilevel,hao2024end,moses2023race,decross2023qubit,sachdeva2024quantum}.   
While variants of QAOA may help improve performance \cite{herrman2022multi,chandarana2022digitized,zhu2022adaptive,blekos2023review}, they often introduce new problems such as more involved classical parameter optimizations or a lack of physical motivation. New quantum algorithms are desirable to address these limitations.

We present a new variational quantum algorithm for combinatorial optimization that achieves high performance with a small number of circuit layers, addressing key limitations of previous approaches.  Our algorithm combines aspects of recent counterdiabatic-inspired ans\"atze \cite{chen2023,chen2024} and variational quantum imaginary time evolution VAR-IT algorithms \cite{mcardle2020quantum,motta2020determining}, along with a variety of algorithmic improvements we developed to reduce the computational resources and improve the convergence rate for combinatorial problems. We tested this approach on a wide variety of instances of the MaxCut problem, which is a standard NP-hard benchmarking problem for quantum and classical algorithms. 

We report several advances obtained from our approach, as verified in its application to MaxCut in theory and simulations.  First, our linear-depth ansatz is theoretically capable of preparing the optimal solution when suitable variational parameters can be identified, significantly improving on the infinite-depth guarantee for QAOA and related approaches.  Second, we demonstrate a new resource-minimal implementation of VAR-IT that consistently finds optimal parameters across a wide variety of problem instances we consider, systematically evolving a fixed initial state to a final optimal solution. Third, we show that quantum entanglement plays an important role the circuit optimization, suggesting the possibility for quantum computational advantages beyond classical simulations representing limited entanglement. Overall, the theoretical and simulated performance addresses central limitations of existing approaches and shows great promise for performant quantum combinatorial optimization. 

We demonstrate success of this approach in circuit optimizations with a truncated version of our ansatz, performed on trapped-ion quantum computers.
We succeed in optimizing ground states of a variety of MaxCut instances, including instances on 3-regular graphs with up to $24$ qubits obtaining fidelities $F \geq 0.5$ on the IonQ Aria computer, while instances with up to $32$ qubits obtain optimal solutions on the IonQ Forte computer, albeit with lower fidelities. These problem instances were chosen completely at random, which demonstrates the robustness of the algorithm in solving any general instance of the MaxCut problem. The quantum circuits used to solve all of these problems were very shallow using only 60 native two qubit ZZ gates. The quantum computational results are among the largest optimizations to date yielding a fidelity $F \geq 0.5$ without any error mitigation or post-processing while using very shallow quantum circuits and only 100 shots to sample the probability distribution from the quantum circuits. These results clearly demonstrate the feasibility of our approach on near term quantum computing hardware, including robustness against shot noise and QPU noise even at 32 qubits. These properties demonstrate clear advantages of this algorithm relative to previous state-of-the-art approaches, paving the way towards larger computations beyond classical simulability.  

\section*{Results}
\subsection*{Combinatorial optimization}

The goal of a combinatorial optimization problem is to minimize a cost function $C(\bm z)$ with an $N$-bit argument $\bm z = (z_1, \ldots, z_N)$.  These problems can often be formulated as quadratic unconstrained binary optimization (QUBO) with $C(\bm z) = \sum_{i<j} J_{i,j} z_i z_j + \sum_i h_i z_i$. 
 For quantum computing these problems are mapped to a diagonal Hamiltonian \cite{lucas2014ising} 
\begin{equation} H_c = \sum_\alpha P_\alpha, \ \ \ H_c\ket{\bm z} = C(\bm z)\ket{\bm z} \end{equation} 
where $P_\alpha$ are Pauli-$Z$ strings and the optimal solution is the ground state of $H_c$. Here we focus on the standard benchmarking problem of weighted MAXCUT. Given a graph $G=(V,E)$ with edge weights $\{w_{i,j}\}$, the goal is to bipartition the vertices such that the summed weight of edges with endpoints in different sets is maximized.  The optimal bipartition $\bm z_\mathrm{opt}$ minimizes $C(\bm z) = \sum_{(i,j) \in E} w_{ij}z_i z_j$ with $z_i \in \{1,-1\}$ (physics convention, $H_c = \sum_{(i,j) \in E} w_{ij}Z_i Z_j$) or $C'(\bm z) = -\sum_{(i,j) \in E} w_{ij}(z_i+z_j-2z_iz_j)/2$ with $z_i \in \{0,1\}$ (computer science convention, $H_c = -\sum_{(i,j) \in E} w_{ij}(1-Z_i Z_j)/2$). Performance is quantified by the approximation ratio
\begin{equation}\label{AR} \mathrm{AR} = \frac{\langle C \rangle}{C_\mathrm{opt}} \end{equation}
where $C_\mathrm{opt}$ is the optimal cost value \cite{farhi2014quantum}. The different conventions are offset by a constant $\sum_{(i,j)\in E} w_{ij}/2$ which affects the approximation ratio.  Following previous work, we use the physics convention for the Sherrington-Kirkpatric model, and the computer science convention for other cases.

\subsection*{Variational Ansatz}

A recent series of papers by Xi Chen and collaborators has considered QAOA-like ansatze that incorporate counterdiabatic terms to improve convergence \cite{chandarana2022digitized,chen2024}, recently culminating in a highly-parameterized ansatz for protein folding problems \cite{chen2023} that we build upon here.
The main ingredients of said ansatz are $ZY$ rotations, which directly rotate between the between the $\pm 1$ sector of $ZZ$ terms of the Hamiltonian.  For MAXCUT, their prescription \cite{chen2024} kept only $ZY$ rotations corresponding to $ZZ$ terms that also occur in $H_c$.  We found that optimization of that ansatz commonly struggled to converge, so we introduce a fully connected ansatz of the form
\begin{align}
\ket{\Psi(\theta)} = \prod_{j = \{(r,q) : r<q\} \in |V|}\exp(i\theta_{j}Z_rY_q)\ket{+}^\otimes
\label{eq:ansatz_form}
\end{align} 
where $j$ indexes all unique pairs of vertices.  The ordering of the vertices is chosen by summed vertex weight as explained in Supplemental Information.
For problems such as MAXCUT consisting of only quadratic $ZZ$ terms, the ground state sector will consist of one or more states locally equivalent (by $X$ flips) to the $N$-qubit GHZ state.  It can be shown that (\ref{eq:ansatz_form}) can reach any such GHZ pair up to inconsequential relative phases, and is overdetermined for such a mapping. 
Each $ZY$ rotation is implementable on quantum hardware in terms of either two CNOTs or one two-qubit gate of the form $\exp(-i\theta ZZ)$ as found on certain trapped ion architectures. The two-qubit gate count for this ansatz is meager: Defining $N_p = N(N-1)/2$, we need only $2N_p$ CNOT gates, or half that if using arbitrary $ZZ$ rotations. In comparison, a single layer of QAOA requires a parameterized two-qubit rotation for each $ZZ$ term in $H_c$, which is less expensive than our ansatze for the least connected graphs and equivalent for fully connected graphs; we will show the cost of our ansatz can be minimized in later sections.

\subsection*{Parameter Optimization} 

System agnostic, classical parameter optimizers are known to suffer from common pitfalls like local minima \cite{anschuetz2022quantum} and vanishing gradients \cite{mcclean2018barren}.  To improve performance we turn to a novel implementation of quantum imaginary time evolution.  Prior implementations of VAR-IT relied on McLachlan's principle \cite{McArdle2019,mcardle2020quantum,alam2023solving}, minimizing 
\begin{align}
    \bigg|\bigg|\left(\frac{\partial}{\partial\tau}+H_c-E_\tau\right)|\Psi(\vec{\theta})\rangle\bigg|\bigg| \label{eq:oldqite}
\end{align}
with respect to angle variations. Variationally evolving in imaginary time on a quantum computer by Eq.~(\ref{eq:oldqite}) requires $N_p(N_p+1)/2$ circuits per iteration, with most relying on controlled ansatz preparation \cite{McArdle2019,mcardle2020quantum}.   
\begin{figure}
    \centering
    \includegraphics[width=.9\columnwidth]{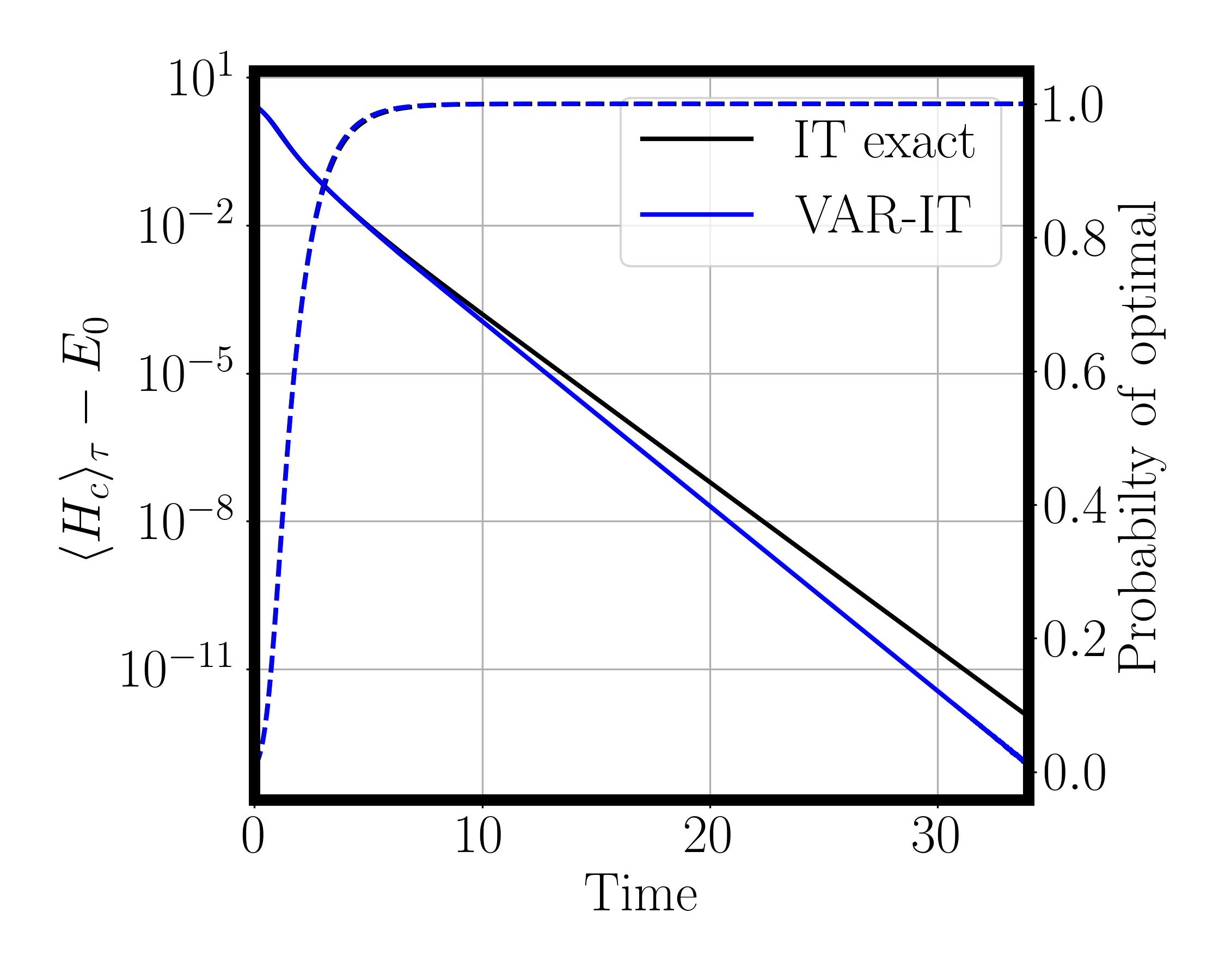}
    \caption{Variational imaginary time evolution, following (\ref{eq:ansatz_form}) and (\ref{eq:varit}), closely matches exact imaginary time evolution for an example 8 vertex, fully connected graph with edge weights sampled from U(0,1) and timestep $\Delta\tau = 0.1 $.}  
    \label{fig:varitvsimtime}
\end{figure}
\begin{figure*}
    \centering
    \includegraphics[width=\textwidth,keepaspectratio]{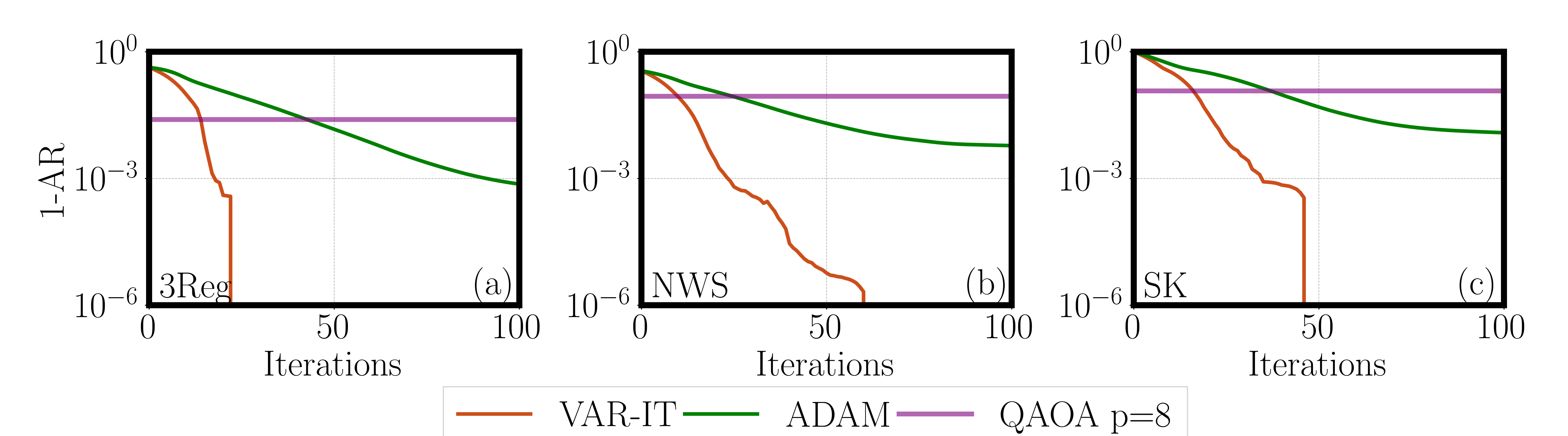}  
    \caption{Simulated average error convergence for 100 random instances of (a) 3 regular, (b) NWS, and (c) SK graphs with 16 vertices. The orange line is VAR-IT, the green line is (\ref{eq:ansatz_form}) optimized with ADAM, and the purple line shows the final optimized error from 8-layer QAOA [from \cite{dupont2024,dupontpersonal}]. }  
    \label{fig:convergence}
\end{figure*}
We instead derive an alternative of VAR-IT that requires fewer function evaluations and no controlled ansatz preparation.  We notice that it is simpler instead to enforce that the expectation value of each $P_\alpha$ appearing in $H_c$ should evolve according the imaginary time as, 
\begin{align}
    \frac{\partial \langle P_{\alpha}\rangle}{\partial\tau} &= \sum_j 2 \mathfrak{R}\left(\langle \Psi(\vec{\theta})|P_\alpha \frac{\partial | \Psi(\vec{\theta})\rangle}{\partial\theta_j}\right)\dot{\theta_j} \notag \\ 
    &=-\langle \Psi(\vec{\theta})|\{P_\alpha,H_c-E_{\tau}\}|\Psi(\vec{\theta})\rangle \,.
\end{align}
Performing these evaluations for each $P_\alpha$ in $H_c$, we arrive at the matrix equation,
\begin{align}
    G\cdot\Dot{\vec{\theta}} = D \label{eq:varit}
\end{align}
where 
\begin{align}
    G_{\alpha,j} = \mathfrak{R}\left(\langle \Psi(\vec{\theta})|P_\alpha \frac{\partial | \Psi(\vec{\theta})\rangle}{\partial\theta_j}\right) \\
    D_{\alpha} = -\frac{1}{2}\langle \Psi(\vec{\theta})|\{P_\alpha,H_c-E_{\tau}\}|\Psi(\vec{\theta})\rangle \,. \label{eq:timeevolutionRHS}
\end{align}
For Ising like Hamiltonians encountered in $H_c$, all $P_\alpha$ commute, so $D$ can be evaluated in a single circuit. Further each column of $G$, corresponding to a given parameter derivative, requires only two circuits if evaluated using the parameter shift rule \cite{mitarai2018quantum}.  To see this note that variations of $P_\alpha$ with respect to a single Pauli rotation [from Eq.~(\ref{eq:ansatz_form})] have the form
\begin{align}
    \langle P_{\alpha} \rangle (\theta_j+\delta) = a_0+b_0\cos(2\delta)+2G_{\alpha,j}\sin(2\delta) \,.
    \label{eq:galpha}
\end{align}
Thus to determine $G$ in QUBO, we need only $2N_p$ circuit evaluations plus one circuit for energy.  Then equation (\ref{eq:varit}) can be inverted numerically to determine $\dot{\vec{\theta}}$ and parameters can be updated with a simple forward Euler step $\vec{\theta} \rightarrow \vec{\theta}+\Delta\tau\dot{\vec{\theta}}$. Fig.~\ref{fig:varitvsimtime} demonstrates the success of our variational imaginary time (VAR-IT) approach for an 8-vertex fully connected MAXCUT problem with uniform random weights sampled from U(0,1) and the ansatz Eq.~(\ref{eq:ansatz_form}).

\subsection*{Improving Convergence} 

For large $N$ graphs, using $H_c$ directly generally imposes harsh requirements on step size as different parameters have rates of change that differ by orders of magnitude.  Additionally, for many families of MAXCUT systems of interest, the spectral gap becomes vanishingly small with large $N$, leading to slow convergence in imaginary-time-based methods.  To mitigate these difficulties, we leverage the manifestly diagonal form of $H_c$, introducing the sigmoidal transformation
\begin{align}
f(H_c) = \left(1+\exp\left(\frac{-\sigma_0 (H_c-E_{\tau})}{4\sigma_{\tau}^2}\right)\right)^{-1}
\end{align}
where $E_{\tau}$, $\sigma_{\tau}$, and $\sigma_{0}$ are the current energy, current standard deviation, and initial standard deviation in the $\ket{+}^{\otimes N}$ state. This function retains the same optimal cut, but stretches the spectrum of states within a standard deviation of the current energy, while pushing farther states towards energies of zero and one. As the imaginary time evolution decreases $E_{\tau}$ and $\sigma_{\tau}$, even dense regions of excited states nearly degenerate with the ground state will approach a gap of 1 in $f(H_c)$.   In the rest of work, Eq.~(\ref{eq:timeevolutionRHS}) is evaluated using $f(H_c)$ and $\langle f(H_c)\rangle $, instead of $H_c$ and $E_\tau$, respectively. 

Numerical and finite shot considerations makes exactly evaluating derivatives in Eq.~(\ref{eq:varit}) a poorly posed task, which manifests most strongly at the end of the imaginary time evolution. We therefore proceed by solving (\ref{eq:varit}) [with $f(H_c)$ in place of $H_c$], discarding the effect of singular values $\lambda_\nu < 0.01\lambda_{max}$ in $G$, where $\lambda_{max}$ is the largest singular value. When cuts of the lowest energy achieve measurement probabilities of 0.5 or larger, we perform Jacobi updates based on a single parameter while holding all the others constant, according to \cite{parrish2019jacobi}; performing this update once for all parameters brings the energy and norm on optimal solutions numerically to one, apart from a single Sherrington-Kirkpatric (SK) instance where VAR-IT converged to a near-degenerate excited state.

\subsection*{Numerical Results} 

We explore three prototypical problem types, 3 regular graphs, random Newman-Watts-Strogatz (NWS) small-world graphs, and the SK spin glass model, with additional details in Supplemental Information \cite{supp}.  These represent linear connectivity, non-trivial clustering type connectivity, and fully connected problem instances, respectively.  In Fig.~\ref{fig:convergence} we show the approximation ratio error $1-$AR from Eq.~(\ref{AR}), produced by our VAR-IT algorithm and by Eq.~(\ref{eq:ansatz_form}) using the ADAM optimizer, averaged over one-hundred 16-vertex instances of the three problem types.  VAR-IT always converges to the optimal energy, unlike the ADAM optimization.  In Supplemental Information we show the fraction of instances that obtain $<1\%$ norm on  the optimal energy after 100 iterations is increasing rapidly with size for the ADAM optimization of NWS and SK graphs \cite{supp};  it is clear that ADAM will suffer a decreasing, and likely vanishing, probability of obtaining the optimal energy as the system size increases, whereas there is no signature of similar failure in VAR-IT in the sizes explored here.  Additionally, VAR-IT converges in significantly fewer iterations than ADAM.  Also plotted are average approximation ratios for $p=8$ QAOA, from Ref.~\cite{dupont2024,dupontpersonal}, which compare unfavorably to both ADAM and VAR-IT. All of the results in Fig.~\ref{fig:convergence} can be improved by the ``relaxed-rounding" postprocessing method of Ref.~\cite{dupont2024}, but still VAR-IT outperforms these leading alternatives, as shown in Supplemental Information \cite{supp}.

\begin{figure}
    \centering
    \includegraphics[width=\columnwidth]{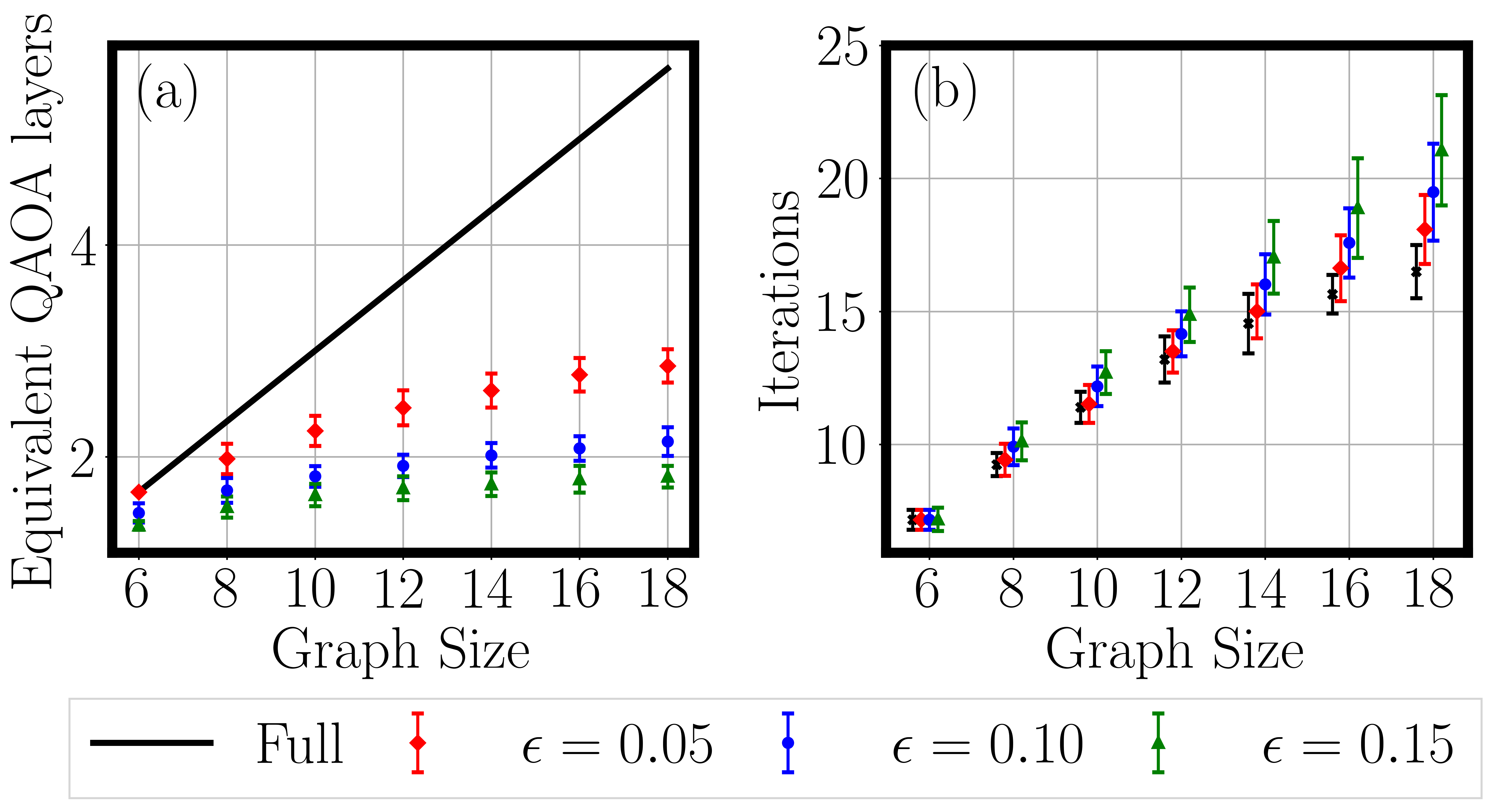}
    \caption{Excluding rotations with $\theta_j < \epsilon$. a) maximum ansatz gate counts  in terms of equivalent layers of QAOA, and b) iterations required to reach an approximation ratio of 1.}
    \label{fig:3reg_adapt}
\end{figure}

When the number of $ZZ$ terms in $H_c$ scales linearly in problem size, QAOA requires $\sim N$ rotation gates while our ansatz requires $\sim N^2$ gates. To combat this shortcoming we explore omitting $ZY$ rotations if angles are smaller than a threshold $\epsilon$ at any point in the optimization.  In Fig. \ref{fig:3reg_adapt}, we apply this adaptive procedure to 100 3-regular graphs for $\epsilon = 0.05,0.10,0.15$.  The maximum number of rotations exploited during each graphs solution in terms of equivalent QAOA layers, is averaged and plotted in the left panel.  For 3-regular graphs and a given $\epsilon$, the circuit cost appears to scale linearly at 2-3 layers of QAOA, with a minor increase in the number of iterations until convergence relative to the full ansatz.  In Supplemental Information, iterations required to converge scale linearly in problem size for SK and NWS, similarly as seen for 3 Regular in Fig. \ref{fig:3reg_adapt}(b).   

\subsection*{Entanglement} 

We argue our algorithm produces large amounts of entanglement, suggesting it will be unsimulatable at scale when using conventional methods representing limited entanglement \cite{orus2014practical,zhou2020limits}.  This is an important prerequisite to potentially obtaining a quantum computational advantage at scale.

\begin{figure}[ht]
    \centering
    \includegraphics[width=\columnwidth]{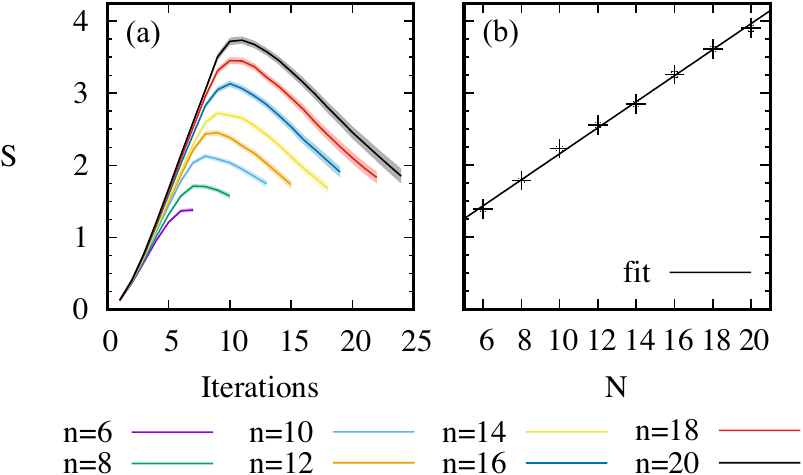}
    \caption{(a) Entanglement entropy $S$ during training of the SK model. (b) Average maximum entropy fit by $S(N) = aN + b$, with $a=0.181 \pm 0.004$ and $b=0.35 \pm 0.05$; the vertical axis is the same as (a). }  
    \label{fig:entropy}
\end{figure}

We analyzed the entanglement entropy $S$ produced by our ansatz across 100 instances of the Sherrington-Kirkpatrick model at various $N$. For each instance we chose a random bipartition into sets of $N/2$ qubits and computed the entanglement entropy of the bipartition for the state produced at each step in the training, with averages over the instances shown in Fig.~\ref{fig:entropy}(a) and colored bands denoting the standard error of the mean.  There is a transition from states with low entanglement at small numbers of iterations, where states are close to the initial product state, to highly entangled states midway through training, and finally to states with decreasing entropy as the training approaches optimal solutions; the curves terminate at the average iteration at which the states reach fidelity $F=1/2$ with the optimal solution subspace, at which point the algorithm switches to Jacobi iterations to isolate the optimal solutions. Fig.~\ref{fig:entropy}(b) shows the maximum entanglement entropy during these training iterations scales as a volume law $S \approx aN +b$. The behavior is similar to what is observed in QAOA \cite{dupont2022entanglement,chen2022much}, except with the crucial difference that we see this transition while {\it training a shallow circuit}, rather than in the dynamics of a deep circuit. Harnessing this entanglement at low depth distinguishes our algorithm as a more practical approach to solving combinatorial problems with a potential advantage.  

\begin{figure*}
    \centering
    \includegraphics[width=\textwidth]{Binary_opt/figs/aria_16q.png}
    \caption{Average and best measured solution value for N=16 on IonQ Aria optimizations with single instance of (a) 3-regular, (b) NWS, and (c) SK model.  }
    \label{fig:aria_runs_16qubits}
\end{figure*}

\subsection*{Experimental Results on IonQ Aria and IonQ Forte}  A few select problem instances of the graph types described in the previous section were chosen to be executed on IonQ quantum processing units (QPUs) Aria and Forte. The IonQ Aria QPU uses up to 25 addressable Ytterbium (Yb) ions linearly arranged in an ion trap, while the IonQ Forte QPU has 36 addressable Ytterbium (Yb) ions. Qubit states are implemented by utilizing two states in the ground hyperfine manifold of the Yb ions. Manipulating the qubits in the Aria and Forte QPUs is done using mode-locked 355nm laser pulses, which drive Raman transitions between the qubit states. By configuring these pulses, arbitrary single-qubit gates and Mølmer-Sørenson type two-qubit gates \cite{sorensen1999quantum} can both be realized. As of this publication, the Aria QPU has demonstrated performance at the level of 25 algorithmic qubits and the Forte QPU has demonstrated performance at the level of 35 algorithmic qubits \cite{IonQ_blog, IonQ_Aria, IonQ_Forte, lubinski2023application}. In order to mitigate the effect of systematic errors on the Aria/Forte QPU, error mitigation via symmetrization is used \cite{maksymov2023enhancing}. After executing multiple circuit variants with distinct qubit to ion mappings, the measurement statistics is aggregated using component-wise averaging.

The graph problem instances chosen to run on IonQ hardware were the 3 regular graphs of sizes 12, 16, 24, 28 and 32 vertices, and the NWS and SK graphs of size 16 vertices. The graph types and sizes were chosen to demonstrate the performance of the VAR-IT algorithm and the IonQ QPU as a function of problem size as well as the capability to handle different types of graphs for a given problem size. The specific graph problem instances were generated completely at random, they were not selected from a set of problem instances for which either the VAR-IT algorithm or the IonQ QPUs show particularly good performance. The 3 regular graphs of sizes 12, 16 and 24 vertices as well as the NWS and SK graphs of size 16 vertices were executed on IonQ Aria while the 3 regular graphs of sizes 28 and 32 vertices were executed on IonQ Forte.

In order to execute the generated graph problem instances on the QPU, the ansatz ~(\ref{eq:ansatz_form}) was truncated to 60 parametrized 2-qubit $ZY$ gates. The truncation was performed taking into account the total QPU execution time for the algorithm and the fact that the ansatz does not necessarily require gates across all pairs of qubits to solve these graph problems. The number of two-qubit gates selected in the truncated ansatz balances the ansatz expressibility, the deleterious effects of QPU noise as more gates are added, and the total execution time on the QPU.

The compiled hardware circuits began with all variational parameters initialized to zero, with the initial state $\ket{+}^{\otimes}$ obtained from Hadamard gates applied to $\ket{0}^\otimes$, and with $60$ native two-qubit gates used in execution of the algorithm. The gradient $G$ in Eq.~(\ref{eq:varit}) was computed using $2\times60=120$ circuit evaluations \cite{mitarai2018quantum}, while the evaluation of $D$ required 1 circuit evaluation, which amounts to 121 circuit evaluations per imaginary time step. All measurements of the quantum circuits were performed in the computational basis with 100 shots per circuit evaluation and measurement.

\begin{figure*}
    \centering
    \includegraphics[width=\textwidth,trim={0 6.5cm 0cm 0},clip]{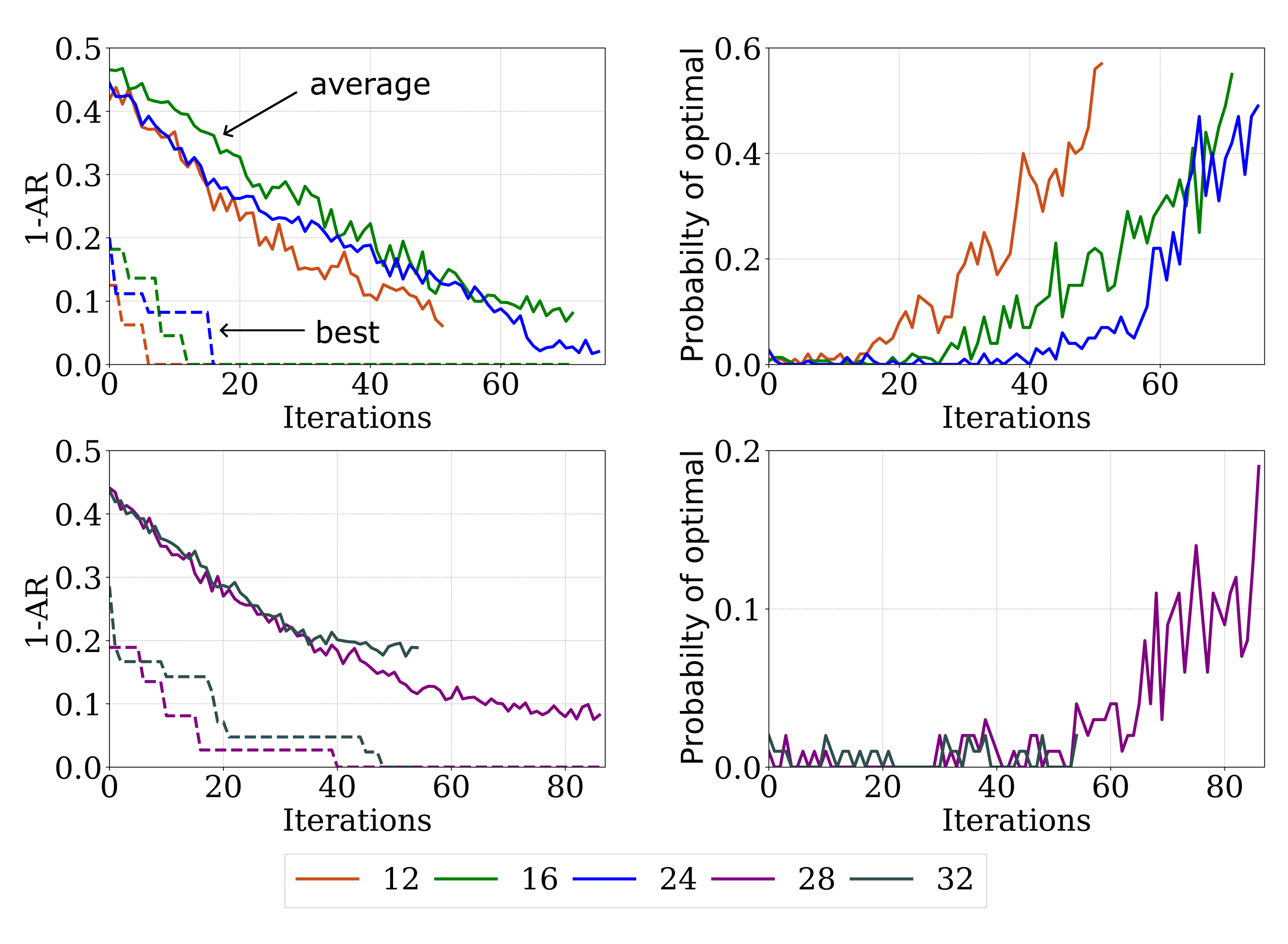}
    \caption{Average error and probability for obtaining the optimal state for various size instances of 3-regular graphs.  Panel (a) and (c) show the average and best sampled solution as a function of iteration obtained on IonQ Aria and Forte respectively.  Panel (b) and (c) contain probability of sampling the optimal solution, on the Aria and Forte respectively.}
    \label{fig:runs_3reg}
\end{figure*}

The hybrid quantum optimization algorithm was executed entirely on the IonQ QPU hardware from initialization until convergence. The evolution of the average error (defined as 1 - Approximation Ratio (AR)) as a function of the iterations of the VAR-IT algorithm is shown in Fig~\ref{fig:aria_runs_16qubits} for three different graphs of 16 vertices - 3 Regular, NWS and SK.  The average error decreases steadily as the higher energy eigenstates of the hamiltonian are projected out and the final distribution peaks on/near the optimal solution. The algorithm converges fairly rapidly to the optimal distribution. This demonstrates the robustness of the VAR-IT algorithm to both shot noise (100 shots per circuit measurement) and QPU noise along with its excellent performance on very different types of graphs with differently weighted edges and connectivity. 

Fig~\ref{fig:runs_3reg} shows the evolution of the average error (solid lines) for 3 Regular graphs of different sizes. The optimizations in Fig~\ref{fig:runs_3reg}(a) were executed on IonQ Aria and those in Fig~\ref{fig:runs_3reg}(c) were executed on IonQ Forte. Once again, the average error is seen to decrease steadily with the algorithm converging close to the optimal distribution within 100 iterations. 

Figs~\ref{fig:aria_runs_16qubits} and ~\ref{fig:runs_3reg} also show error of the best solution sampled in the 100-shot evaluation of $\langle H_c\rangle$ at a given iteration (dashed lines). These converge rapidly to 0.0 for all graph problem instances indicating that an optimal solution has been sampled. The execution of the algorithm on the QPUs was stopped once the optimal solution was sampled and the probability of sampling the optimal solution was boosted to an adequately high value or time limitations prevented further execution.

Fig~\ref{fig:runs_3reg}(b) and (c) show the probability of measuring any (degenerate) optimal solution as a function of the iterations. The probability of the optimal solution is very small during the first few iterations and then starts to increase rapidly as the algorithm begins to converge, ultimately reaching large optimal solution probabilities of $\gtrsim 0.5$ with 24 or fewer qubits and $0.19$ with 28 qubits (execution of the 32-qubit instance was terminated early due to time constraints, but this still succeeded in sampling the optimal solution). The 28 and 32 qubits 3 regular graph problem instances represent some of the largest full hybrid optimization ever performed on quantum hardware to date, to the best of the authors' knowledge. Furthermore, the achievement of fidelity $F \geq 0.5$ with over 20 qubits, from device optimization and without error mitigation or post-processing, is a significant achievement enabled by this advanced hardware and algorithm implementation.

\section{Conclusion and Outlook} 

We have developed an algorithm for quantum combinatorial optimization that makes several important advances over leading alternatives, in terms of optimization procedure and iterations, convergence behaviors, and circuit depth.  Key features of our approach include counterdiabatic Hamiltonian generators capable of preparing optimal solutions, training through VAR-IT with a nonlinear objective that emphasizes the ground state, and resource-efficient generation of volume-law entanglement during training. Widespread success of the algorithm was evident in numerical simulations of diverse problem instances with up to 20 qubits, which consistently converged to optimal solutions, as well as algorithm execution on quantum computers with up to 32 qubits.

We have demonstrated full hybrid on-QPU optimization of the MaxCut problem for graphs of varying types, connectivity and sizes. Our results have shown that a truncated version of our algorithm is capable of finding optimal solutions with 32 or fewer qubits, with convergence to high optimal solution probability. The problem instances were chosen completely at random which suggests the robustness of the algorithm in solving generic instances of the MaxCut problem, as anticipated from the numerical simulations. The high quality results also demonstrated the robustness against shot noise and QPU noise even at 32 qubits. These properties demonstrate clear advantages of our approach relative to the previous state-of-the-art.

In future work, we expect the combination of our algorithm with problem decomposition techniques \cite{karypis1998fast,METIS1998,ushijima2021multilevel,acharya2024decomposition,ponce2023graph, maciejewski2024multilevel} will enable hierarchical solutions to combinatorial problems that begin to scale towards industrially-relevant sizes. Another important frontier is the extension of our approach to higher-order binary optimization problems where classical methods fail to return optimal solutions for problems with $\lesssim 100$ variables \cite{shaydulin2023evidence}, where optimal solution convergence as we observed here could potentially discover better solutions than any conventional method. The low depth, minimal resource requirements, high entanglement, superior empirical performance, and scalability of our approach make it a promising candidate for solving these and other interesting problems in the future.

\section{Acknowledgements} 

We thank Joseph Wang, Eugene Dumitrescu, Jim Ostrowski, Ryan Bennink, George Siopsis, Rizwanul Alam, Noah Bauer, and Rebekah Herrman for discussions of variational quantum optimization algorithms.  We thank Maxime Dupont for discussion and sharing numerical results from Ref.~\cite{dupont2024}. We thank Ruslan Shaydulin for discussion and for bringing Refs.~\cite{hao2024end,moses2023race,decross2023qubit,sachdeva2024quantum} to our attention. T.D.M was supported by the U.S. Department of Energy, Office of Science, National Quantum Information Science Research Centers, Quantum Science Center and devised the ansatz, optimization algorithm, performed the simulations, and contributed to writing. P.C.L.~was supported by the DARPA ONISQ program under award W911NF-20-20051 and contributed to design, analysis, writing, and entanglement computations. This research used resources of the Compute and Data Environment for Science (CADES) at the Oak Ridge National Laboratory, which is supported by the Office of Science of the U.S. Department of Energy under Contract No. DE-AC05-00OR22725.

\bibliographystyle{apsrev4-1}
\bibliography{references.bib}

\pagebreak
\widetext
\begin{center}
\textbf{\large Supplemental Materials: Performant near-term quantum combinatorial optimization}
\end{center}
\setcounter{equation}{0}
\setcounter{figure}{0}
\setcounter{table}{0}
\setcounter{page}{1}
\makeatletter
\renewcommand{\theequation}{S\arabic{equation}}
\renewcommand{\thefigure}{S\arabic{figure}}
\renewcommand{\bibnumfmt}[1]{[S#1]}

\section{Ansatz}
In pursuit of a solution of the binary optimization problem, we perform an analysis of the QAOA ansatz,  
\begin{align}
\ket{\vec{\theta},\vec{\gamma}}_p = \prod_{j=0}^{p}\exp(-i\gamma_j H_m)\exp(-i\theta_j H_c)\ket{+}^\otimes ,
\end{align}
where $H_m$ is the standard mixing Hamiltonian and $H_c$ is the cost Hamiltonian with an eigenspectrum $H_c \ket{\bm z} = C(\bm z)\ket{\bm z}$ that contains the set of classical solution values 
\begin{align}
H_m = -\sum_{j \in V}X_j \\
H_c = \sum_{j,k \in E}\omega_{jk}Z_j Z_k = \sum_\alpha \lambda_\alpha P_\alpha \,.
\end{align}

In the limit as $p \rightarrow \infty$, and with appropriately chosen $\vec{\theta}$ and $\vec{\gamma}$, the QAOA ansatz yields adiabatic evolution from the ground state $\ket{+}^{\otimes n}$ of $H_m$ to the optimal-solution ground state $\ket{\bm z_\mathrm{opt}}$ of $H_c$. At finite $p$ that are implementable on current quantum computers, QAOA has been shown to be successful at approximating $\ket{\bm z_\mathrm{opt}}$ by variationally optimizing the parameters $p$. Analyzing the QAOA/adiabatic ansatz by combining the two sequential applications of a single layer under its equivalent single exponential is illustrative.

The leading order term after the sum of the driving and mixing Hamiltonians is their single commutator
\begin{align}
-\frac{\gamma_j*\theta_j}{2}[H_m,H_c] = -\frac{i\gamma_j*\theta_j}{2}\sum_{k,l \in E}\omega_{kl}(Y_k Z_l+Z_k Y_l) \,.
\end{align}
Previous work in accelerating convergence within QAOA found these are precisely the terms most important for accelerating convergence \cite{chandarana2022,Zhu2022}.  

Is it clear from the following relation
\begin{align}
\exp(\frac{i\pi ZY}{2})
\left\{
    \begin{array}{lr}
        \ket{00}\pm \ket{11} \\
        \ket{01}\pm \ket{10} 
    \end{array}
\right\} = 
\left\{
    \begin{array}{lr}
        \mp \ket{01}\mp \ket{10} \\
         \quad   \ket{00}\pm \ket{11} 
    \end{array}
\right\}  
\end{align}
that $ZY$ rotations exactly rotate norm between the $\pm 1$ sector of $ZZ$ terms of the Hamiltonian.  Based on these findings, we employ a parameterized ansatz of the form 
\begin{align}
\ket{\Psi(\theta)} = \prod_{j = \{(r,q) : r<q\} \in |V|}\exp(i\theta_{j}Z_rY_q)\ket{+}^\otimes
\end{align} 
where $j$ indexes all unique pairs of vertices.  
The ordering is important for robust convergence, so we introduce the quantity 
\begin{align}
\rho_j = \sum_{k \in E}|\omega_{jk}|
\end{align}
which represents the relative importance of each vertex in the overall $H_c$. Thus in this work, we order the $ZY$ rotations in the ansatz by their corresponding size in $\rho$.  For a vertex set $|V|$ indexed now by their order in $\rho$, the ansatz exploited in this work has an explicit form
\begin{align}
\ket{\Psi(\theta)} = \exp(i\theta_{j}Z_{N-1}Y_{N})\ldots \exp(i\theta_{j}Z_0Y_2)\exp(i\theta_{j}Z_0Y_1)\ket{+}^\otimes\,.
\end{align} 
\begin{figure*}[ht]
    \centering
    \includegraphics[width=\textwidth]{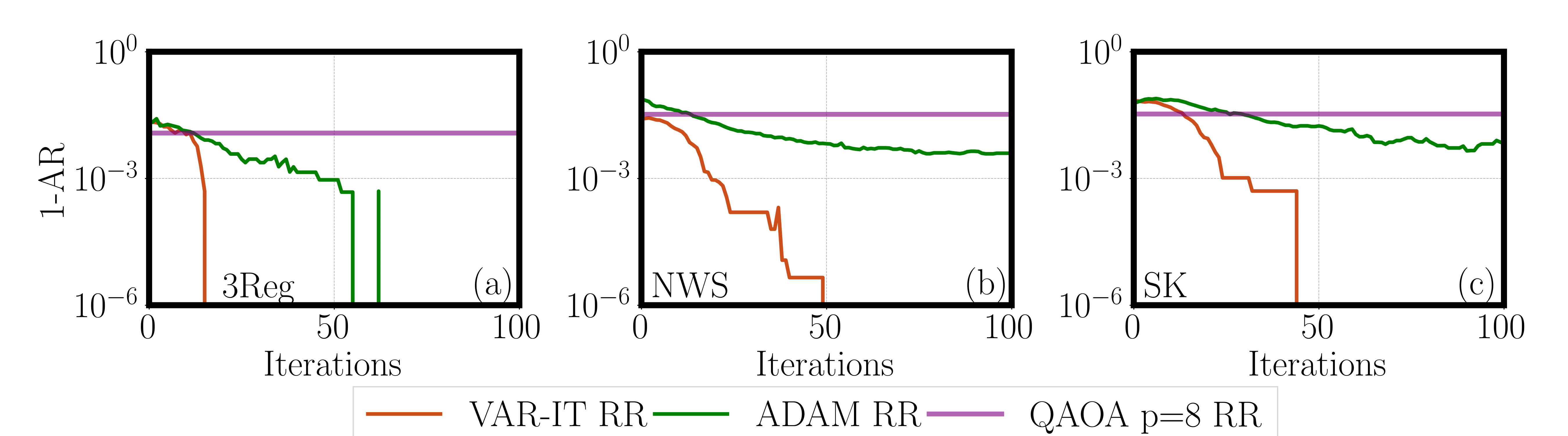}  
    \caption{Average error convergence for 100 random instances of (a) 3 regular, (b) NWS, and (c) SK graphs with 16 vertices and relaxed rounding performed. The orange line is VAR-IT, the green line is (\ref{eq:ansatz_form}) optimized with ADAM, and the purple line shows the optimized error from 8-layer QAOA [from \cite{dupont2024,dupontpersonal}]. }  
    \label{fig:convergence_rr}
\end{figure*}
\section{Problem types}
The different problem types explored in this work are dictated by the $\omega_{ij}$ between vertices, commonly referred to as the adjacency matrix.  All graphs in this work are generated by the Python package NetworkX \cite{networkx}.

{\it Sherrington-Kirkpatrick (SK)} -- The SK model in this work is defined as the fully connected graph type, with $\omega_{ij}$ chosen between $\pm1$ chosen with equal probability. 

{\it Newman-Watts-Strogatz small-world graphs (NWS)} -- In order to create models of this type, vertices of a ring graph are connected with its $k$ nearest neighbors.  In this work, $k=4$.  Afterwards, for each resulting edge (i,j), in the ring with connections to k nearest neighbors, add a new edge (i,l) with probability $p$, in this work $p=0.5$, to a randomly-chosen vertex $l$. Then $\omega_{ij}$ are assigned randomly in the interval $(0,1]$.

{\it Random 3 Regular (3Reg)} -- This model arises from graphs where each vertex is connected to three other vertices randomly.  $\omega_{ij}$ are all assigned a uniform value of 1 in this case.

\section{Relaxed Rounding}
In \cite{dupont2024}, the authors introduced a quantum analogue of the classical relax and round method.  They showed that it can easily be used to augment wavefunction methods like QAOA, and always improves the approximation ratio. It relies on the establishing the correlation matrix $\chi_{ij}$, defined as 
\begin{align}
\chi_{ij} = (\delta_{ij}-1)\langle Z_iZ_j\rangle \,.
\end{align}
Then $\chi$ can be diagonalized classically to yield its eigenvectors $\{\alpha_\nu \in \mathbb{R}^N\}$.  These can be translated to their closest bitstrings by rounding $\{z_\nu \leftarrow sign(\alpha_\nu) \in {\pm 1}^N\}$. Then the $z_\nu$ which minimizes the $C(z)$ is chosen as the outcome of this quantum,relax, and round procedure.  We apply this procedure at each iteration of our optimization procedure and report the findings in Fig. \ref{fig:convergence_rr}.

\section{Further Numerical Results}
In numerical results section of this work, we presented the error of the approximation ratio when using our VAR-IT method for optimization of Eq. \ref{eq:ansatz_form}, and the ADAM optimizer.  

\begin{figure}[h]
    \centering
    \includegraphics[width=\columnwidth]{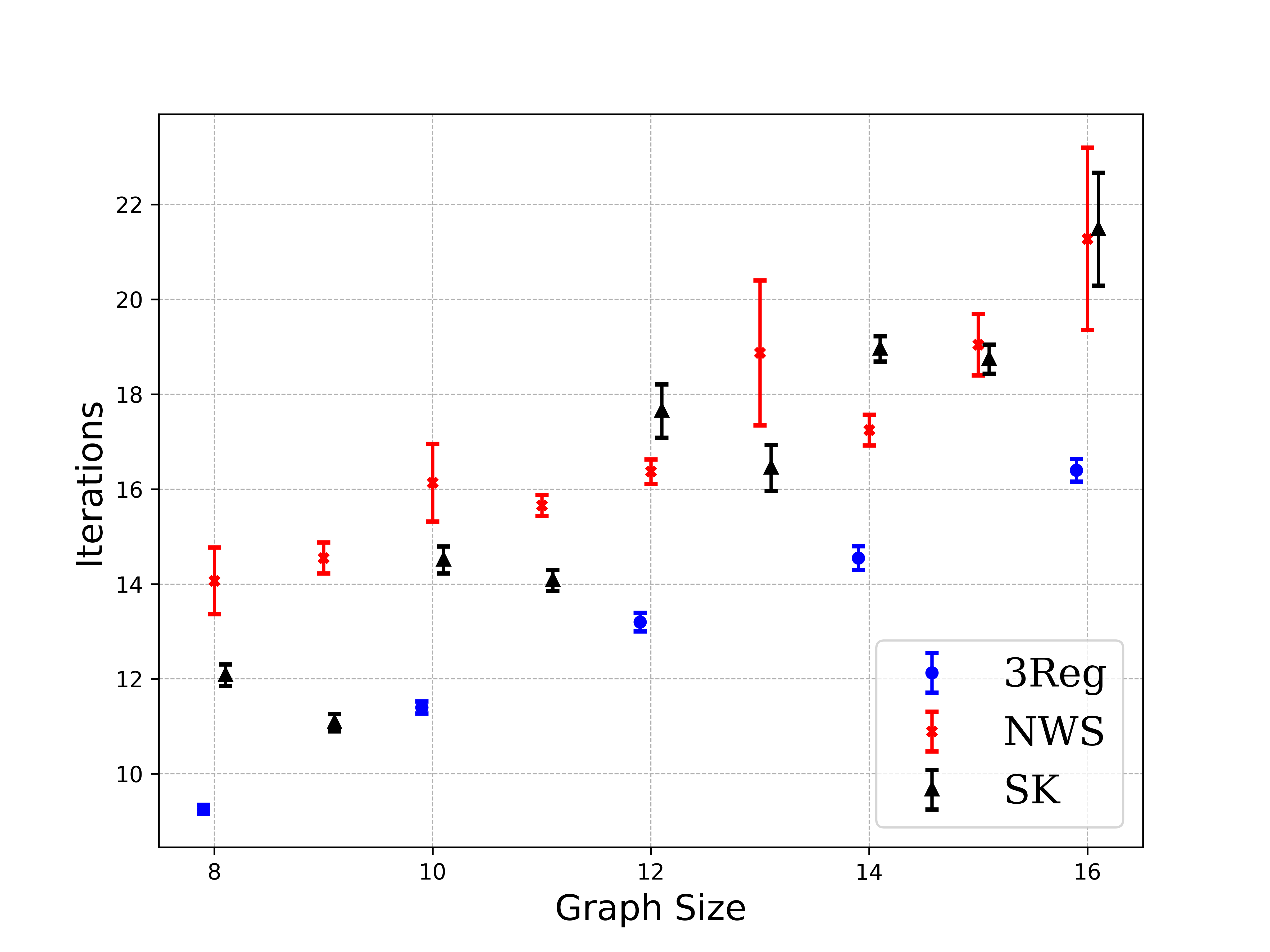}  
    \caption{ Average iterations required as a function of graph size. }  
    \label{fig:iterationtoconv}
\end{figure}
\begin{figure*}[!b]
    \centering
    \includegraphics[width=.9\textwidth]{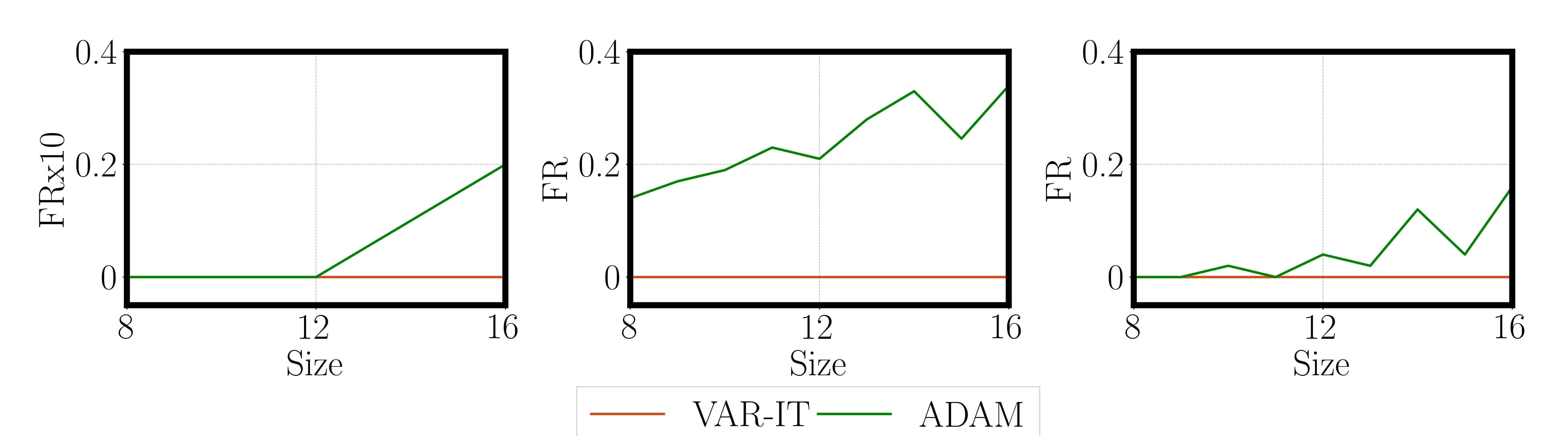}  
    \caption{Fraction of instances with $<1\%$ norm on optimal solutions after 100 iterations. }  
    \label{fig:failure}
\end{figure*}
In addition to the VAR-IT method always converging to an optimal solution, Fig. \ref{fig:iterationtoconv} demonstrates that for the sizes explored in this work, the iterations required to reach convergence for all three problem types explored in this work appear to be scaling linearly in problem size.  Additionally, Fig. \ref{fig:failure} shows that an ansatz optimized with ADAM will yield increasing failures as problem size grows, while there does not appear to be any such signature with the VAR-IT optimization procedure.


\end{document}


\begin{myhideenv}
\title{Performant near-term quantum combinatorial optimization}
\author{Titus Morris}
\affiliation{Quantum Information Science Section, Oak Ridge National Laboratory, Oak Ridge, TN 37831, USA}

\author{Phillip C. Lotshaw}
\affiliation{Quantum Information Science Section, Oak Ridge National Laboratory, Oak Ridge, TN 37831, USA}
\blfootnote{This manuscript has been authored by UT-Battelle, LLC under Contract No. DE-AC05-00OR22725 with the U.S. Department of Energy. The United States Government retains and the publisher, by accepting the article for publication, acknowledges that the United States Government retains a non-exclusive, paid-up, irrevocable, world-wide license to publish or reproduce the published form of this manuscript, or allow others to do so, for United States Government purposes. The Department of Energy will provide public access to these results of federally sponsored research in accordance with the DOE Public Access Plan (http://energy.gov/downloads/doe-public-access-plan).}

\date{\today}

\begin{abstract}
We present a variational quantum algorithm for solving combinatorial optimization problems with linear-depth circuits.  Our algorithm uses an ansatz composed of Hamiltonian generators designed to control each term in the target combinatorial function, along with parameter updates following a modified version of quantum imaginary time evolution. We evaluate this ansatz in numerical simulations that target solutions to the MAXCUT problem. The state evolution is shown to closely mimic imaginary time evolution, and its optimal-solution convergence is further improved using adaptive transformations of the classical Hamiltonian spectrum, while resources are minimized by pruning optimized gates that are close to the identity.  With these innovations, the algorithm consistently converges to optimal solutions, with interesting highly-entangled dynamics along the way. This performant and resource-minimal approach is a promising candidate for potential quantum computational advantages on near-term quantum computing hardware. 
\end{abstract}

\maketitle

{\it Introduction} -- Combinatorial optimization is widely viewed as a scientific computing challenge where quantum computers may make a significant impact \cite{preskill2018quantum,herman2023quantum,abbas2023quantum}. Research on this topic has been motivated by several encouraging results obtained for the quantum approximate optimization algorithm (QAOA) \cite{farhi2014quantum,hogg2000quantum} as well as its variants \cite{hadfield2019quantum,herman2023constrained,ushijima2021multilevel,tate2023warm,bartschi2020grover,wurtz2021classically}.  These include theoretical analyses \cite{wurtz2021MAXCUT,diez2023quantum,wurtz2022counterdiabaticity}, which have shown high QAOA approximation ratios 
for the SK model \cite{farhi2022quantum} and for Maximum-cut on large girth graphs \cite{basso2021quantum}, as well as a variety of encouraging results on hardware \cite{ebadi2022quantum,maciejewski2023design,harrigan2021quantum,pagano2020quantum,niroula2022constrained,shaydulin2023qaoa,pelofske2023quantum} and numerical simulations \cite{zhou2020quantum,lotshaw2021empirical,herrman2021impact,golden2023numerical,crooks2018performance}, such as a numerical scaling advantage for the low-autocorrelation binary sequence (LABS) problem \cite{shaydulin2023evidence}.  Modified QAOA-like ans\"atze have also determined ways to improve performance by introducing additional parameterized gates \cite{herrman2022multi}, including gates with counter-diabatic Hamiltonians \cite{chandarana2022digitized} or based on Hamiltonians selected to maximize the energy convergence rate \cite{zhu2022adaptive}, as well as many others \cite{blekos2023review}. 

Despite encouraging progress there are several drawbacks to current QAOA-based approaches.  The primary limitation is that deep circuits are required for high performance \cite{farhi2022quantum,lykov2023sampling,lotshaw2023approximate,marwaha2021local} while noise at these depths places severe limitations on theoretical and experimental performance \cite{stilck2021limitations,lotshaw2022scaling,weidenfeller2022scaling,de2023limitations}. Graph instances can even be designed so that any given QAOA depth $p$ cannot outperform classical algorithms in the noiseless case \cite{bravyi2020obstacles}. One reason large depths are required is that a single QAOA layer only entangles qubits that are adjacent in a given problem graph; hence the algorithm can require significant numbers of layers to build correlations across all qubits \cite{farhi2020quantum,farhi2020quantum2}, and further layers to remove this entanglement and approach the optimal classical solution \cite{dupont2022entanglement}.  It has been argued that entanglement actually serves as a barrier to QAOA performance due to the prohibitive depths required to remove it \cite{chen2022much}.  This is concerning---entanglement should be helping our quantum algorithms, not hurting them. There have been approaches to heuristically modify the QAOA ansatz to reduce the depth or improve performance, such as  Refs.~\cite{herrman2022multi,chandarana2022digitized,zhu2022adaptive,blekos2023review}.  While these and related approaches do give noteworthy empirical improvements, they often do not improve on the theoretical $p \to \infty$ guarantee of standard QAOA, and can also introduce their own problems in terms of more involved parameter optimizations, as well as losing QAOA's physical motivation in terms of an approximation to an optimal annealing algorithm \cite{brady2021optimal,brady2021behavior}. It is desirable to find other physically motivated approaches that do not have these shortcomings.

In this letter, we present a new algorithm for quantum combinatorial optimization that overcomes several issues of previous approaches such as QAOA. Building on recent work on counter-diabatic inspired ans\"atze \cite{chen2023,chen2024}, we show that our algorithm is guaranteed to be able to prepare the optimal solution at linear depth given suitable parameters.  We use parameters that are optimized through a systematic procedure, based on variational quantum imaginary time evolution and with an energy-dependent nonlinear objective that isolates the optimal solutions during training.  Numerical simulations show that the final states at different training iterations transition from approximate product states, to volume-law entangled states, and to final states confined to the optimal solution subspace. Our ansatz accesses these highly entangled states during training of short-depth circuits, rather than in deep circuits, as in QAOA. 
Furthermore, we find that a small number of iterations succeed in identifying optimal solutions to a wide variety of problems we consider. The high-performance, low-depth, and modest training costs distinguish our algorithm as a prime candidate for successful near-term quantum optimization. 

{\it Combinatorial optimization} -- The goal of a combinatorial optimization problem is to minimize a cost function $C(\bm z)$ with an $N$-bit argument $\bm z = (z_1, \ldots, z_N)$.  These problems can often be formulated as quadratic unconstrained binary optimization (QUBO) with $C(\bm z) = \sum_{i<j} J_{i,j} z_i z_j + \sum_i h_i z_i$. 
 For quantum computing these problems are mapped to a diagonal Hamiltonian \cite{lucas2014ising} 
\begin{equation} H_c = \sum_\alpha P_\alpha, \ \ \ H_c\ket{\bm z} = C(\bm z)\ket{\bm z} \end{equation} 
where $P_\alpha$ are Pauli-$Z$ strings and the optimal solution is the ground state of $H_c$. Here we focus on the standard benchmarking problem of weighted MAXCUT. Given a graph $G=(V,E)$ with edge weights $\{w_{i,j}\}$, the goal is to bipartition the vertices such that the summed weight of edges with endpoints in different sets is maximized.  The optimal bipartition $\bm z_\mathrm{opt}$ minimizes $C(\bm z) = \sum_{(i,j) \in E} w_{ij}z_i z_j$ with $z_i \in \{1,-1\}$ (physics convention, $H_c = \sum_{(i,j) \in E} w_{ij}Z_i Z_j$) or $C'(\bm z) = -\sum_{(i,j) \in E} w_{ij}(z_i+z_j-2z_iz_j)/2$ with $z_i \in \{0,1\}$ (computer science convention, $H_c = -\sum_{(i,j) \in E} w_{ij}(1-Z_i Z_j)/2$). Performance is quantified by the approximation ratio
\begin{equation}\mathrm{AR} = \frac{\langle C \rangle}{C_\mathrm{opt}} \end{equation}
where $C_\mathrm{opt}$ is the optimal cost value. The different conventions are offset by a constant $\sum_{(i,j)\in E} w_{ij}/2$ which affects the approximation ratio.  Following previous work, we use the physics convention for the Sherrington-Kirkpatric model, and the computer science convention for other cases.

{\it Variational Ansatz} -- A recent series of papers by Xi Chen and collaborators has considered QAOA-like ansatze that incorporate counterdiabatic terms to improve convergence \cite{chandarana2022digitized,chen2024}, recently culminating in a highly-parameterized ansatz for protein folding problems \cite{chen2023} that we build upon here.
The main ingredients of said ansatz are $ZY$ rotations, which directly rotate between the between the $\pm 1$ sector of $ZZ$ terms of the Hamiltonian.  For MAXCUT, their prescription \cite{chen2024} kept only $ZY$ rotations corresponding to $ZZ$ terms that also occur in $H_c$.  We found that optimization of that ansatz commonly struggled to converge, so we introduce a fully connected ansatz of the form
\begin{align}
\ket{\Psi(\theta)} = \prod_{j = \{(r,q) : r<q\} \in |V|}\exp(i\theta_{j}Z_rY_q)\ket{+}^\otimes
\label{eq:ansatz_form}
\end{align} 
where $j$ indexes all unique pairs of vertices.  The ordering of the vertices is chosen by summed vertex weight as explained in S1.
For problems such as MAXCUT consisting of only quadratic $ZZ$ terms, the ground state sector will consist of one or more states locally equivalent (by $X$ flips) to the $N$-qubit GHZ state.  It can be shown that (\ref{eq:ansatz_form}) can reach any such GHZ pair up to inconsequential relative phases, and is overdetermined for such a mapping. 
Each $ZY$ rotation is implementable on quantum hardware in terms of either two CNOTs or one two-qubit gate of the form $\exp(-i\theta ZZ)$ as found on certain trapped ion architectures such as Quantinuum devices. The two-qubit gate count for this ansatz is meager: Defining $N_p = N(N-1)/2$, we need only $2N_p$ CNOT gates, or half that if using arbitrary $ZZ$ rotations. In comparison, a single layer of QAOA requires a parameterized two-qubit rotation for each $ZZ$ term in $H_c$, which is less expensive than our ansatze for the least connected graphs and equivalent for fully connected graphs; we will show the cost of our ansatz can be minimized in later sections.

{\it Parameter Optimization} -- System agnostic, classical parameter optimizers are known to suffer from common pitfalls like local minima \cite{anschuetz2022quantum} and vanishing gradients \cite{mcclean2018barren}.  To improve performance we turn to a novel implementation of quantum imaginary time (QITE).  Prior implementations of QITE relied on McLachlan's principle \cite{McArdle2019,mcardle2020quantum,alam2023solving}, minimizing 
\begin{align}
    \bigg|\bigg|\left(\frac{\partial}{\partial\tau}+H_c-E_\tau\right)|\Psi(\vec{\theta})\rangle\bigg|\bigg| \label{eq:oldqite}
\end{align}
with respect to angle variations. Variationally evolving in imaginary time on a quantum computer by Eq. \ref{eq:oldqite} requires $N_p(N_p+1)/2$ circuits per iteration, with most relying on controlled ansatz preparation \cite{McArdle2019,mcardle2020quantum}.   
\begin{figure}[H]
    \centering
    \includegraphics[width=.7\columnwidth]{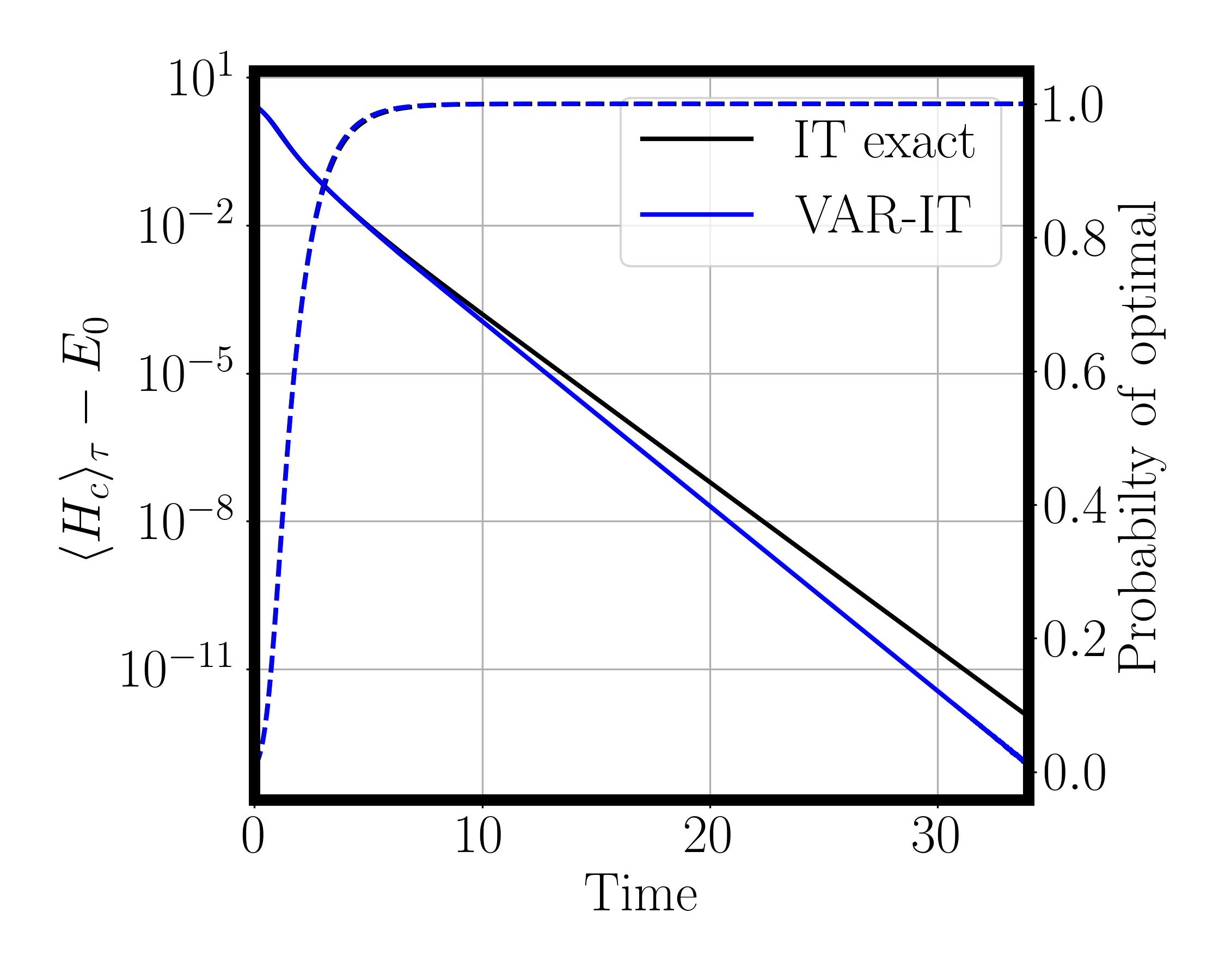}
    \caption{Variational imaginary time evolution, following (\ref{eq:ansatz_form}) and (\ref{eq:varit}), closely matches exact imaginary time evolution for an example 8 vertex, fully connected graph with edge weights sampled from U(0,1) and timestep $\Delta\tau = 0.1 $.}  
    \label{fig:varitvsimtime}
\end{figure}
We instead derive an alternative of QITE that requires fewer function evaluations and no controlled ansatz preparation.  We notice that it is simpler instead to enforce that the expectation value of each $P_\alpha$ appearing in $H_c$ should evolve according the imaginary time as, 
\begin{align}
    \frac{\partial \langle P_{\alpha}\rangle}{\partial\tau} &= \sum_j 2 \mathfrak{R}\left(\langle \Psi(\vec{\theta})|P_\alpha \frac{\partial | \Psi(\vec{\theta})\rangle}{\partial\theta_j}\right)\dot{\theta_j} \notag \\ 
    &=-\langle \Psi(\vec{\theta})|\{P_\alpha,H_c-E_{\tau}\}|\Psi(\vec{\theta})\rangle \,.
\end{align}
Performing these evaluations for each $P_\alpha$ in $H_c$, we arrive at the matrix equation,
\begin{align}
    G\cdot\Dot{\vec{\theta}} = D \label{eq:varit}
\end{align}
where 
\begin{align}
    G_{\alpha,j} = \mathfrak{R}\left(\langle \Psi(\vec{\theta})|P_\alpha \frac{\partial | \Psi(\vec{\theta})\rangle}{\partial\theta_j}\right) \\
    D_{\alpha} = -\frac{1}{2}\langle \Psi(\vec{\theta})|\{P_\alpha,H_c-E_{\tau}\}|\Psi(\vec{\theta})\rangle \,. \label{eq:timeevolutionRHS}
\end{align}
For Ising like Hamiltonians encountered in $H_c$, all $P_\alpha$ commute, so $D$ can be evaluated in a single circuit. Further each column of $G$, corresponding to a given parameter derivative, requires only two circuits.  To see this note that variations of $P_\alpha$ with respect to a single Pauli rotation [from Eq.~(\ref{eq:ansatz_form})] have the form
\begin{align}
    \langle P_{\alpha} \rangle (\theta_j+\delta) = a_0+b_0\cos(2\delta)+2G_{\alpha,j}\sin(2\delta) \,.
    \label{eq:galpha}
\end{align}
Thus to determine $G$ in QUBO, we need only $2N_p$ circuit evaluations plus one circuit for energy.  Then equation (\ref{eq:varit}) can be inverted numerically to determine $\dot{\vec{\theta}}$ and parameters can be updated with a simple forward Euler step $\vec{\theta} \rightarrow \vec{\theta}+\Delta\tau\dot{\vec{\theta}}$. Fig.~\ref{fig:varitvsimtime} demonstrates the success of our variational imaginary time (VAR-IT) approach for an 8-vertex fully connected MAXCUT problem with uniform random weights sampled from U(0,1) and the ansatz Eq.~(\ref{eq:ansatz_form}).

{\it Improving Convergence} -- For large $N$ graphs, using $H_c$ directly generally imposes harsh requirements on step size as different parameters have rates of change that differ by orders of magnitude.  Additionally, for many families of MAXCUT systems of interest, the spectral gap becomes vanishingly small with large $N$, leading to slow convergence in imaginary-time-based methods.  To mitigate these difficulties, we leverage the manifestly diagonal form of $H_c$, introducing the sigmoidal transformation
\begin{align}
f(H_c) = \left(1+\exp\left(\frac{-\sigma_0 (H_c-E_{\tau})}{4\sigma_{\tau}^2}\right)\right)^{-1}
\end{align}
where $E_{\tau}$, $\sigma_{\tau}$, and $\sigma_{0}$ are the current energy, current standard deviation, and initial standard deviation in the $\ket{+}^{\otimes N}$ state. This function retains the same optimal cut, but stretches the spectrum of states within a standard deviation of the current energy, while pushing farther states towards energies of zero and one. As the imaginary time evolution decreases $E_{\tau}$ and $\sigma_{\tau}$, even dense regions of excited states nearly degenerate with the ground state will approach a gap of 1 in $f(H_c)$.   In the rest of work, Eq.~(\ref{eq:timeevolutionRHS}) is evaluated using $f(H_c)$ and $\langle f(H_c)\rangle $, instead of $H_c$ and $E_\tau$, respectively. 

\begin{figure*}[ht]
    \centering
    \includegraphics[width=\textwidth]{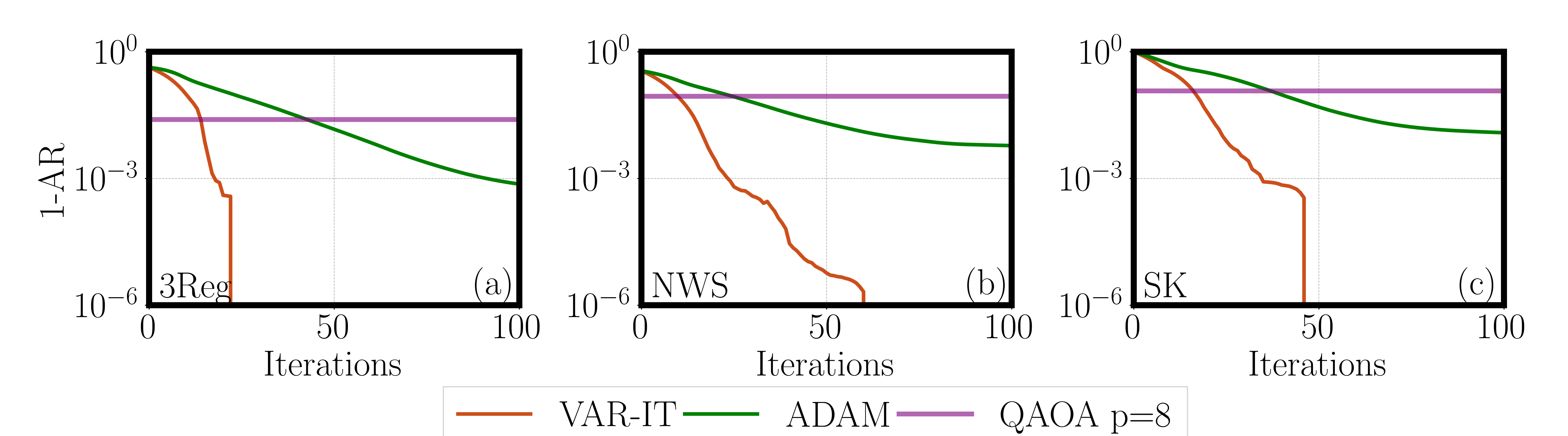}  
    \caption{Average error convergence for 100 random instances of (a) 3 regular, (b) NWS, and (c) SK graphs with 16 vertices. The orange line is VAR-IT, the green line is (\ref{eq:ansatz_form}) optimized with ADAM, and the purple line shows the optimized error from 8-layer QAOA [from \cite{dupont2024,dupontpersonal}]. }  
    \label{fig:convergence}
\end{figure*}

Numerical and finite shot considerations makes exactly evaluating derivatives in Eq.~(\ref{eq:varit}) a poorly posed task, which manifests most strongly at the end of the imaginary time evolution. We therefor proceed by solving (\ref{eq:varit}) [with $f(H_c)$ in place of $H_c$] until the lowest energy cut measured achieves half the measurements, discarding the effect of singular values $\lambda_\nu < 0.01\lambda_{max}$ in $G$, where $\lambda_{max}$ is the largest singular value. Once this is achieved, we perform Jacobi updates based on a single parameter while holding all the others constant, according to \cite{parrish2019jacobi}; performing this update once for all parameters brings the energy and norm on optimal solutions numerically to one, apart from a single Sherrington-Kirkpatric (SK) instance where VAR-IT converged to a near-degenerate excited state.

{\it Numerical Results} -- We explore three prototypical problem types, 3 regular graphs, random Newman-Watts-Strogatz (NWS) small-world graphs, and the SK spin glass model with more details in Supplemental Information \cite{supp}.  These represent linear connectivity, non-trivial clustering type connectivity, and fully connected problem instances, respectively.  In Fig.~\ref{fig:convergence} we show the approximation ratio error produced by our VAR-IT algorithm and by Eq.~(\ref{eq:ansatz_form}) using the ADAM optimizer, averaged over one-hundred 16-vertex instances of the three problem types.  VAR-IT always converges to the optimal energy, unlike the ADAM optimization.  In Supplemental Information we show the fraction of instances that obtain $<1\%$ norm on  the optimal energy after 100 iterations is increasing rapidly with size for the ADAM optimization of NWS and SK graphs \cite{supp};  it is clear that ADAM will suffer a decreasing, and likely vanishing, probability of obtaining the optimal energy as the system size increases, whereas there is no signature of similar failure in VAR-IT in the sizes explored here.  Additionally, VAR-IT converges in significantly fewer iterations than ADAM.  Also plotted are average approximation ratios for $p=8$ QAOA, from Ref.~\cite{dupont2024,dupontpersonal}, which compare unfavorably to both ADAM and VAR-IT. All of the results in Fig.~\ref{fig:convergence} can be improved by the ``relaxed-rounding" postprocessing method of Ref.~\cite{dupont2024}, but still VAR-IT outperforms these leading alternatives, as shown in Supplemental Information \cite{supp}.

\begin{figure}
    \centering
    \includegraphics[width=\columnwidth]{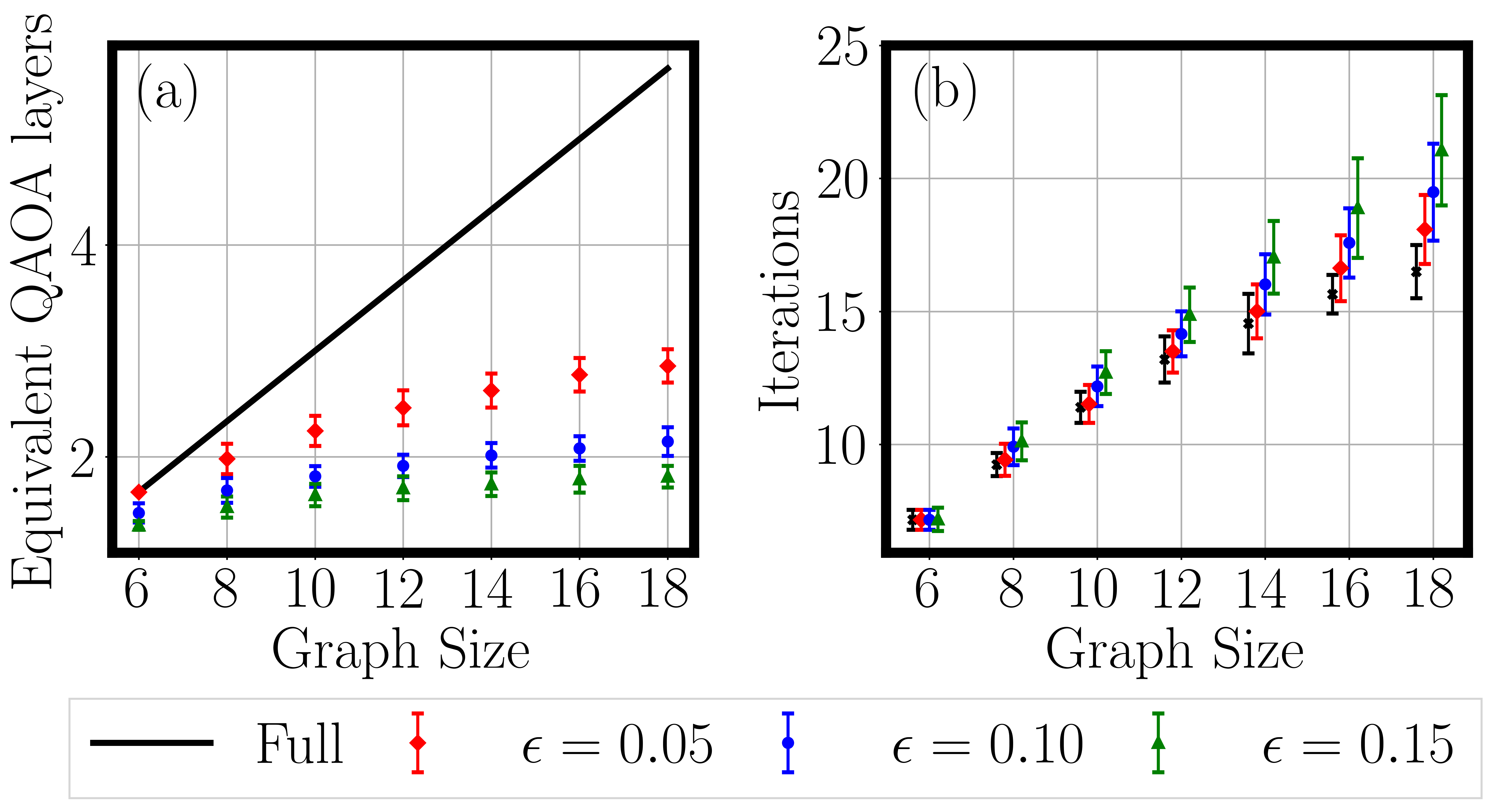}
    \caption{Excluding rotations with $\theta_j < \epsilon$. a) maximum ansatz gate counts  in terms of equivalent layers of QAOA, and b) iterations required to reach an approximation ratio of 1.}
    \label{fig:3reg_adapt}
\end{figure}
When the number of $ZZ$ terms in $H_c$ scales linearly in problem size, QAOA requires $\sim N$ rotation gates while our ansatz requires $\sim N^2$ gates. To combat this shortcoming we explore omitting $ZY$ rotations if angles are smaller than a threshold $\epsilon$ at any point in the optimization.  In Fig. \ref{fig:3reg_adapt}, we apply this adaptive procedure to 100 3-regular graphs for $\epsilon = 0.05,0.10,0.15$.  The maximum number of rotations exploited during each graphs solution in terms of equivalent QAOA layers, is averaged and plotted in the left panel.  For 3-regular graphs and a given $\epsilon$, the circuit cost appears to scale linearly at 2-3 layers of QAOA, with a minor increase in the number of iterations until convergence relative to the full ansatz.  In Supplemental Information, iterations required to converge scale linearly in problem size for SK and NWS, similarly as seen for 3 Regular in Fig. \ref{fig:3reg_adapt}(b).

{\it Entanglement} --  
If our algorithm produces large amounts of entanglement then it suggests the algorithm cannot be simulated at scale and therefor has potential for a quantum advantage, while low amounts of entanglement would seem to preclude quantum advantage due to simulability by tensor networks or related methods \cite{zhou2020limits}.

We analyzed the entanglement entropy $S$ produced by our ansatz across 100 instances of the Sherrington-Kirkpatrick model at various $N$. For each instance we chose a random bipartition into sets of $N/2$ qubits and computed the entanglement entropy of the bipartition for the state produced at each step in the training, with averages over the instances shown in Fig.~\ref{fig:entropy}(a) and colored bands denoting the standard error of the mean.  There is a transition from states with low entanglement at small numbers of iterations, where states are close to the initial product state, to highly entangled states midway through training, and finally to states with decreasing entropy as the training approaches optimal solutions; the curves terminate at the average iteration at which the states reach fidelity $F=1/2$ with the optimal solution subspace, at which point the algorithm switches to Jacobi iterations to isolate the optimal solutions. Fig.~\ref{fig:entropy}(b) shows the maximum entanglement entropy during these training iterations scales as a volume law $S \approx aN +b$. The behavior is similar to what is observed in QAOA \cite{dupont2022entanglement,chen2022much}, except with the crucial difference that we see this transition while {\it training a shallow circuit}, rather than in the dynamics of a deep circuit. Harnessing this entanglement at low depth distinguishes our algorithm as a more practical approach to solving combinatorial problems with a potential advantage.  

\begin{figure}[ht]
    \centering
    \includegraphics[width=\columnwidth]{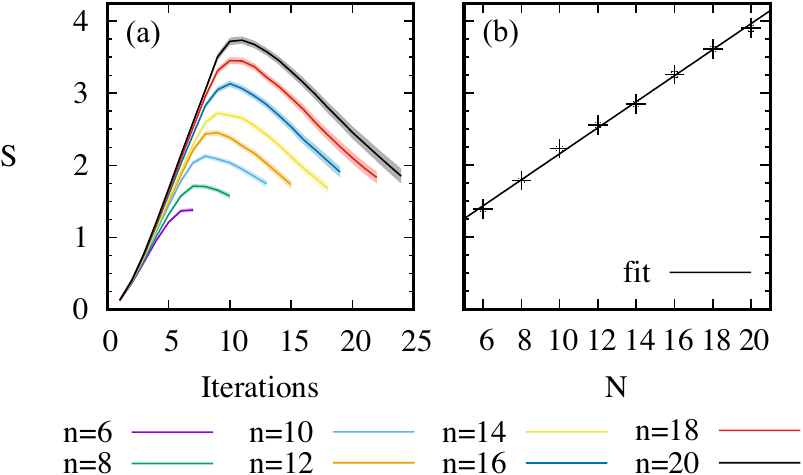}
    \caption{(a) Entanglement entropy $S$ during training of the SK model. (b) Average maximum entropy fit by $S(N) = aN + b$, with $a=0.181 \pm 0.004$ and $b=0.35 \pm 0.05$; the vertical axis is the same as (a). }  
    \label{fig:entropy}
\end{figure}

{\it Conclusion} -- We developed an algorithm for quantum combinatorial optimization that makes several important advances over leading alternatives, in terms of optimization procedure and iterations, convergence behaviors, and circuit depth; widespread success of the algorithm is evident in simulations of diverse problems with up to 20 qubits.  Key features of our approach include counterdiabatic Hamiltonian generators capable of preparing optimal solutions, training through QITE with a nonlinear objective that emphasizes the ground state, and resource-efficient generation of volume-law entanglement during training.  We expect the low depth and minimal training requirements of our algorithm to enable quantum computers to solve combinatorial optimization problems with several tens of qubits in the immediate future, with extensions to larger sizes and scaling analyses beyond classical simulability as quantum computers continue to develop.  The low depth, minimal resource requirements, high entanglement, and superior empirical performance make our approach a leading contender for near-term quantum computational advantage in combinatorial optimization.

{\it Acknowledgements} -- We thank Joseph Wang, Eugene Dumitrescu, Jim Ostrowski, Ryan Bennink, George Siopsis, Rizwanul Alum, Noah Bauer, and Rebekah Herrman for discussions of variational quantum optimization algorithms.  We thank Maxime Dupont for discussion and sharing numerical results from Ref.~\cite{dupont2024}. T.D.M was supported by the U.S. Department of Energy, Office of Science, National Quantum Information Science Research Centers, Quantum Science Center and devised the ansatz, optimization algorithm, performed the simulations, and contributed to writing. P.C.L.~was supported by the DARPA ONISQ program under award W911NF-20-20051 and contributed to design, analysis, writing, and entanglement computations. This research used resources of the Compute and Data Environment for Science (CADES) at the Oak Ridge National Laboratory, which is supported by the Office of Science of the U.S. Department of Energy under Contract No. DE-AC05-00OR22725.

\bibliographystyle{apsrev4-1}
\bibliography{references.bib}
\end{myhideenv}

\pagebreak
\widetext
\begin{center}
\textbf{\large Supplemental Materials: Performant near-term quantum combinatorial optimization}
\end{center}
\setcounter{equation}{0}
\setcounter{figure}{0}
\setcounter{table}{0}
\setcounter{page}{1}
\makeatletter
\renewcommand{\theequation}{S\arabic{equation}}
\renewcommand{\thefigure}{S\arabic{figure}}
\renewcommand{\bibnumfmt}[1]{[S#1]}

\section{Ansatz}
In pursuit of a solution of the binary optimization problem, we perform an analysis of the QAOA ansatz,  
\begin{align}
\ket{\vec{\theta},\vec{\gamma}}_p = \prod_{j=0}^{p}\exp(-i\gamma_j H_m)\exp(-i\theta_j H_c)\ket{+}^\otimes ,
\end{align}
where $H_m$ is the standard mixing Hamiltonian and $H_c$ is the cost Hamiltonian with an eigenspectrum $H_c \ket{\bm z} = C(\bm z)\ket{\bm z}$ that contains the set of classical solution values 
\begin{align}
H_m = -\sum_{j \in V}X_j \\
H_c = \sum_{j,k \in E}\omega_{jk}Z_j Z_k = \sum_\alpha \lambda_\alpha P_\alpha \,.
\end{align}

In the limit as $p \rightarrow \infty$, and with appropriately chosen $\vec{\theta}$ and $\vec{\gamma}$, the QAOA ansatz yields adiabatic evolution from the ground state $\ket{+}^{\otimes n}$ of $H_m$ to the optimal-solution ground state $\ket{\bm z_\mathrm{opt}}$ of $H_c$. At finite $p$ that are implementable on current quantum computers, QAOA has been shown to be successful at approximating $\ket{\bm z_\mathrm{opt}}$ by variationally optimizing the parameters $p$. Analyzing the QAOA/adiabatic ansatz by combining the two sequential applications of a single layer under its equivalent single exponential is illustrative.

The leading order term after the sum of the driving and mixing Hamiltonians is their single commutator
\begin{align}
-\frac{\gamma_j*\theta_j}{2}[H_m,H_c] = -\frac{i\gamma_j*\theta_j}{2}\sum_{k,l \in E}\omega_{kl}(Y_k Z_l+Z_k Y_l) \,.
\end{align}
Previous work in accelerating convergence within QAOA found these are precisely the terms most important for accelerating convergence \cite{chandarana2022,Zhu2022}.  

Is it clear from the following relation
\begin{align}
\exp(\frac{i\pi ZY}{2})
\left\{
    \begin{array}{lr}
        \ket{00}\pm \ket{11} \\
        \ket{01}\pm \ket{10} 
    \end{array}
\right\} = 
\left\{
    \begin{array}{lr}
        \mp \ket{01}\mp \ket{10} \\
         \quad   \ket{00}\pm \ket{11} 
    \end{array}
\right\}  
\end{align}
that $ZY$ rotations exactly rotate norm between the $\pm 1$ sector of $ZZ$ terms of the Hamiltonian.  Based on these findings, we employ a parameterized ansatz of the form 
\begin{align}
\ket{\Psi(\theta)} = \prod_{j = \{(r,q) : r<q\} \in |V|}\exp(i\theta_{j}Z_rY_q)\ket{+}^\otimes
\end{align} 
where $j$ indexes all unique pairs of vertices.  
The ordering is important for robust convergence, so we introduce the quantity 
\begin{align}
\rho_j = \sum_{k \in E}|\omega_{jk}|
\end{align}
which represents the relative importance of each vertex in the overall $H_c$. Thus in this work, we order the $ZY$ rotations in the ansatz by their corresponding size in $\rho$.  For a vertex set $|V|$ indexed now by their order in $\rho$, the ansatz exploited in this work has an explicit form
\begin{align}
\ket{\Psi(\theta)} = \exp(i\theta_{j}Z_{N-1}Y_{N})\ldots \exp(i\theta_{j}Z_0Y_2)\exp(i\theta_{j}Z_0Y_1)\ket{+}^\otimes\,.
\end{align} 
\begin{figure*}[ht]
    \centering
    \includegraphics[width=\textwidth]{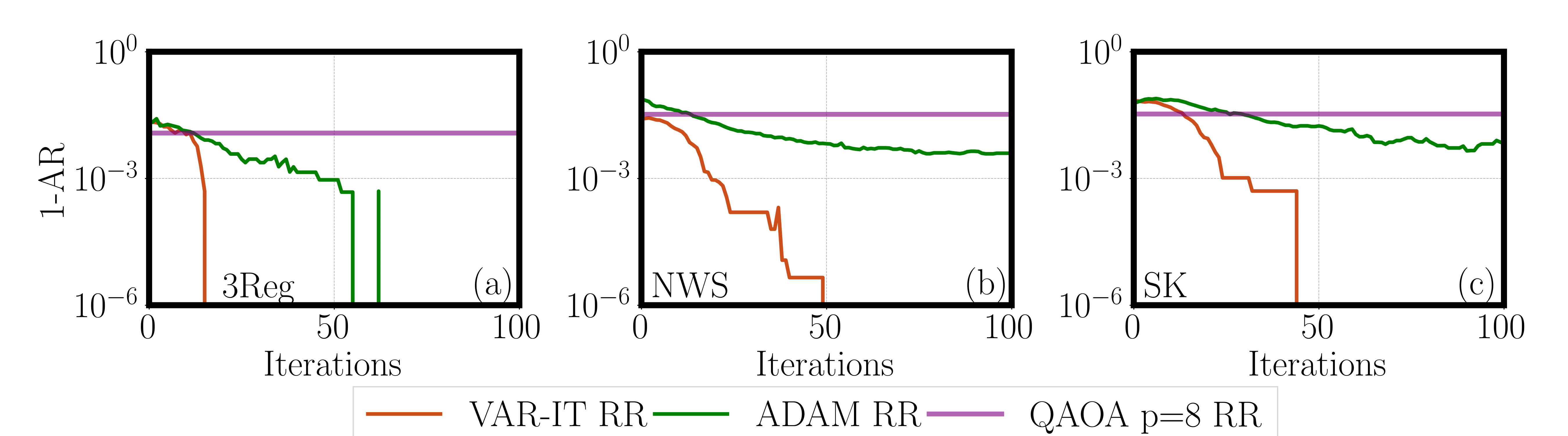}  
    \caption{Average error convergence for 100 random instances of (a) 3 regular, (b) NWS, and (c) SK graphs with 16 vertices and relaxed rounding performed. The orange line is VAR-IT, the green line is (\ref{eq:ansatz_form}) optimized with ADAM, and the purple line shows the optimized error from 8-layer QAOA [from \cite{dupont2024,dupontpersonal}]. }  
    \label{fig:convergence_rr}
\end{figure*}
\section{Problem types}
The different problem types explored in this work are dictated by the $\omega_{ij}$ between vertices, commonly referred to as the adjacency matrix.  All graphs in this work are generated by the Python package NetworkX \cite{networkx}.

{\it Sherrington-Kirkpatrick (SK)} -- The SK model in this work is defined as the fully connected graph type, with $\omega_{ij}$ chosen between $\pm1$ chosen with equal probability. 

{\it Newman-Watts-Strogatz small-world graphs (NWS)} -- In order to create models of this type, vertices of a ring graph are connected with its $k$ nearest neighbors.  In this work, $k=4$.  Afterwards, for each resulting edge (i,j), in the ring with connections to k nearest neighbors, add a new edge (i,l) with probability $p$, in this work $p=0.5$, to a randomly-chosen vertex $l$. Then $\omega_{ij}$ are assigned randomly in the interval $(0,1]$.

{\it Random 3 Regular (3Reg)} -- This model arises from graphs where each vertex is connected to three other vertices randomly.  $\omega_{ij}$ are all assigned a uniform value of 1 in this case.

\section{Relaxed Rounding}
In \cite{dupont2024}, the authors introduced a quantum analogue of the classical relax and round method.  They showed that it can easily be used to augment wavefunction methods like QAOA, and always improves the approximation ratio. It relies on the establishing the correlation matrix $\chi_{ij}$, defined as 
\begin{align}
\chi_{ij} = (\delta_{ij}-1)\langle Z_iZ_j\rangle \,.
\end{align}
Then $\chi$ can be diagonalized classically to yield its eigenvectors $\{\alpha_\nu \in \mathbb{R}^N\}$.  These can be translated to their closest bitstrings by rounding $\{z_\nu \leftarrow sign(\alpha_\nu) \in {\pm 1}^N\}$. Then the $z_\nu$ which minimizes the $C(z)$ is chosen as the outcome of this quantum,relax, and round procedure.  We apply this procedure at each iteration of our optimization procedure and report the findings in Fig. \ref{fig:convergence_rr}.

\section{Further Numerical Results}
In numerical results section of this work, we presented the error of the approximation ratio when using our VAR-IT method for optimization of Eq. \ref{eq:ansatz_form}, and the ADAM optimizer.  

\begin{figure}[h]
    \centering
    \includegraphics[width=\columnwidth]{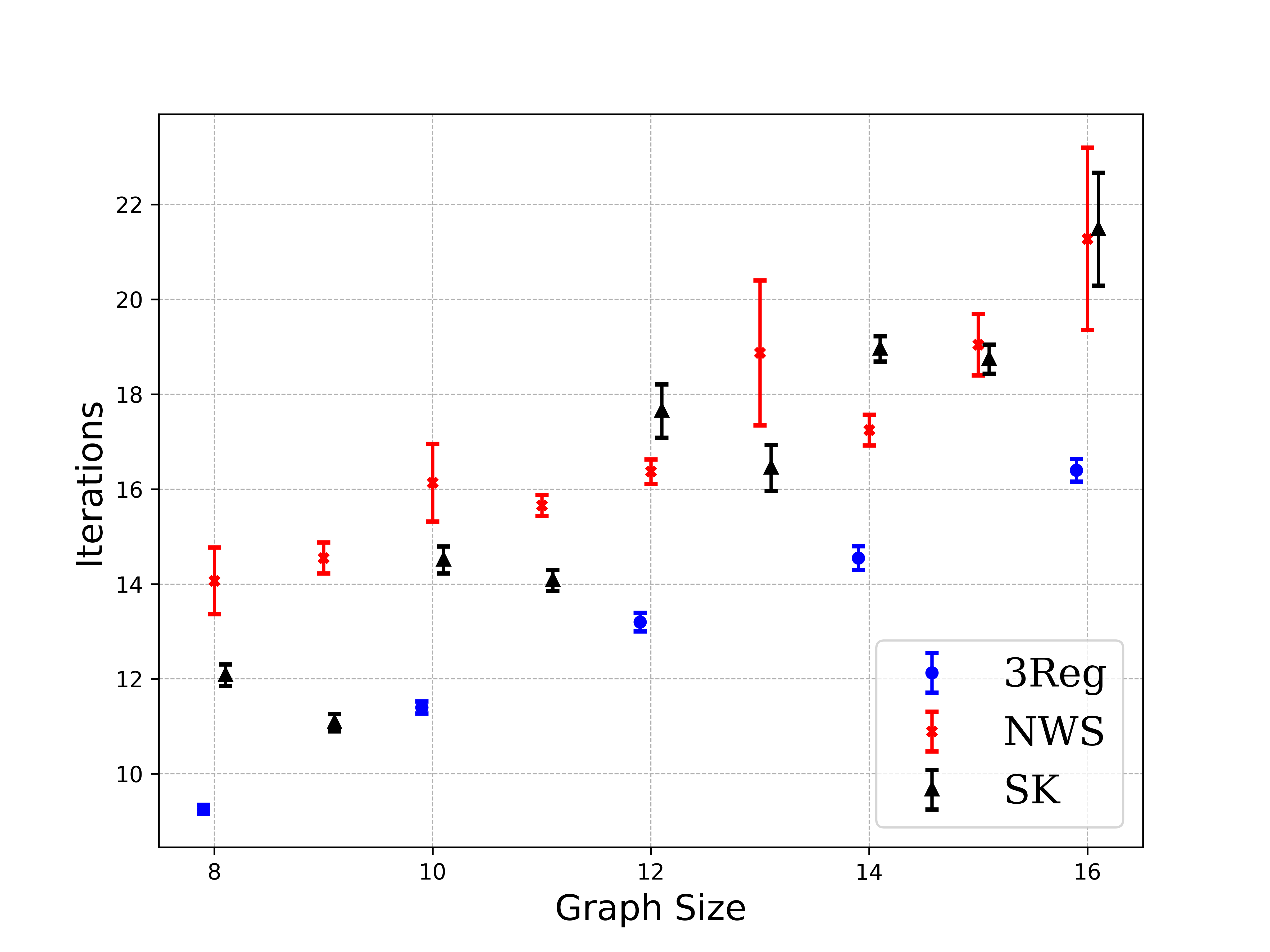}  
    \caption{ Average iterations required as a function of graph size. }  
    \label{fig:iterationtoconv}
\end{figure}
\begin{figure*}[!b]
    \centering
    \includegraphics[width=.9\textwidth]{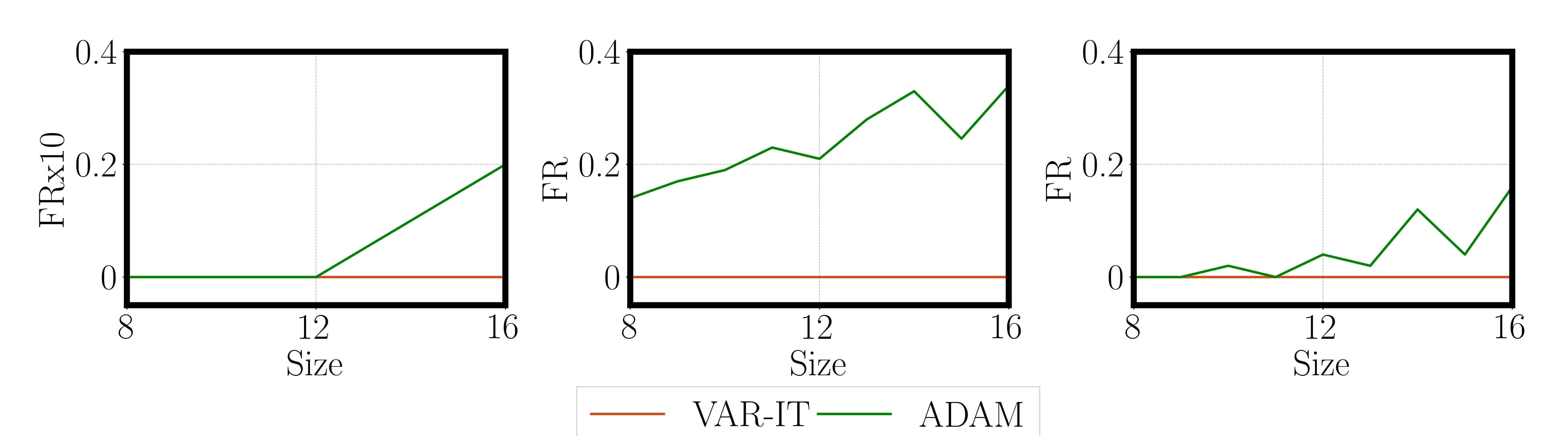}  
    \caption{Fraction of instances with $<1\%$ norm on optimal solutions after 100 iterations. }  
    \label{fig:failure}
\end{figure*}
In addition to the VAR-IT method always converging to an optimal solution, Fig. \ref{fig:iterationtoconv} demonstrates that for the sizes explored in this work, the iterations required to reach convergence for all three problem types explored in this work appear to be scaling linearly in problem size.  Additionally, Fig. \ref{fig:failure} shows that an ansatz optimized with ADAM will yield increasing failures as problem size grows, while there does not appear to be any such signature with the VAR-IT optimization procedure.
